\documentclass[11pt,a4paper]{article}
\usepackage{graphicx}
\usepackage{epstopdf}
\usepackage{amsmath,amssymb,amsbsy,color}
\usepackage{multirow}
\usepackage{hyperref}
\usepackage[ruled,vlined,noline]{algorithm2e}
\usepackage{algorithmic}
\usepackage[numbers,sort&compress]{natbib}
\def\lsim{\raise0.3ex\hbox{$<$\kern-0.75em\raise-1.1ex\hbox{$\sim$}}}
\def\gsim{\raise0.3ex\hbox{$>$\kern-0.75em\raise-1.1ex\hbox{$\sim$}}}
\def\simgt{\rlap{\lower 3.5 pt\hbox{$\mathchar \sim$}}\raise 2.0pt \hbox {$>$}}
\def\simlt{\rlap{\lower 3.5 pt\hbox{$\mathchar \sim$}}\raise 2.0pt \hbox {$<$}}

\begin{document}
\begin{flushright}
  \small{
    UTHEP-732, UTCCS-P-122
  }
\end{flushright}
\vspace{2mm}

\begin{center}
  {\Large\bf
    Hadronic vacuum polarization contribution to the muon $g-2$\\ with 2+1 flavor lattice QCD on a larger than (10 fm$)^4$ lattice at the physical point
  }
\end{center}
\vspace{5mm}
\begin{center}
  Eigo~Shintani$^{a}$ and
  Yoshinobu~Kuramashi$^{b}$\\
  (PACS Collaboration)
\\[4mm]
{\small\it
  $^a$RIKEN Center for Computational Science, Kobe, Hyogo 650-0047, Japan}
\\
{\small\it
  $^b$Center for Computational Sciences, University of Tsukuba, Tsukuba, Ibaraki 305-8577, Japan}
\\[10mm]
\end{center}
\begin{abstract}
  We study systematic uncertainties in the lattice QCD computation of hadronic vacuum polarization (HVP) contribution to the muon $g-2$. We investigate three systematic effects;  finite volume (FV) effect, cutoff effect, and integration scheme dependence. We evaluate the FV effect at the physical pion mass on two different volumes of (5.4 fm$)^4$ and (10.8 fm$)^4$ using the PACS10 configurations at the same cutoff scale. For the cutoff effect, we compare two types of lattice vector operators, which are local and conserved (point-splitting) currents, by varying the cutoff scale on a larger than (10 fm$)^4$ lattice at the physical point. For the integration scheme dependence, we compare the results between the coordinate- and momentum-space integration schemes at the physical point on a (10.8 fm$)^4$ lattice. Our result for the HVP contribution to the muon $g-2$ is given by $a_\mu^{\rm hvp} = 737(9)(^{+13}_{-18})\times 10^{-10}$ in the continuum limit, where the first error is statistical and the second one is systematic.

\end{abstract}

\section{Introduction}
The muon anomalous magnetic moment $(g-2)_\mu$ has been a key observable for a proof of predictability of quantum field theory. We expect that there might be a sign of the new physics beyond SM (BSM) in the muon $g-2$ anomaly, which is 3$\sigma$ to 4$\sigma$ deviation between the standard model (SM) prediction and the BNL experiment~\cite{Bennett:2004pv,Bennett:2006fi} suggested in 2004. In order to establish that the $(g-2)_\mu$ experiments in FermiLab and J-PARC~\cite{Flay:2016vuw,Shimomura:2015aza} aiming at a factor of 4 to 5 improvement from the BNL experiment is forthcoming. However, the high precision experiments are not sufficient for the search of the BSM physics~\cite{Lindner:2016bgg} since the magnitude of theoretical uncertainty in the SM prediction has not been comparable to that in the new experiments yet. The biggest uncertainty left in the SM prediction is coming from the hadronic vacuum polarization (HVP) effect, which is the leading order of the hadronic contribution to $(g-2)_\mu$ denoted by $a_\mu^{\rm hvp}$. The phenomenological estimate of $a_\mu^{\rm hvp}$~\cite{Gourdin:1969dm,Davier:2010nc,Hagiwara:2011af,Jegerlehner:2015stw,Davier:2017zfy,Keshavarzi:2018mgv}, which has been employed in the SM prediction, is obtained by the integrated hadronic R-ratio measured in $e^+e^-$ annihilation experiment. Including several hadronic decay channels with a particular choice of the center-of-mass energy $\sqrt s$ window, in which the perturbative QCD is used for $\sqrt s\simeq 2$ GeV, $a_\mu^{\rm hvp}$ is phenomenologically estimated at a 0.4\% level of precision~\cite{Olive:2016xmw}.

Lattice QCD (LQCD) is another approach to estimate $a_\mu^{\rm hvp}$ totally independent of the phenomenological estimate. This is a theoretical calculation based on the first principle of QCD, whereas the current precision of LQCD estimate is roughly an order of magnitude lower than the phenomenological one, and it then does not satisfy accuracy to search the BSM physics (see a recent review \cite{Meyer:2018til} and references therein). The main difficulty of the LQCD calculation is that, in the Euclidean space-time, the detailed behavior of the HVP contribution with high precision is required around the peak position of the QED kernel, which is significantly below the hadronic scale of $\Lambda_{\rm QCD}$. In such a low-energy region, corresponding to a long distance in the coordinate space-time, it is not an easy task to make a high precision measurement of the HVP contribution because of the exponentially diminishing signal-to-noise ratio in a deeply infrared regime. In addition, the contribution of the $\rho$ resonance state decreases in this regime, while two-pion or three-pion state contributions, which are possible decay modes of the vector resonance, become prominent. This means that a sufficiently large volume at the physical point, where the $\rho$ resonance has an open threshold and the multipion states are allowed, is required in the LQCD calculation to correctly estimate the HVP contribution. Furthermore, it is imperative for LQCD to take account of the cutoff effect to obtain $a_\mu^{\rm hvp}$ in the continuum limit. So the LQCD determination of $a_\mu^{\rm hvp}$ at a subpercent precision is still a challenging task.

Recent LQCD calculations~\cite{Chakraborty:2016mwy,DellaMorte:2017dyu,Borsanyi:2017zdw,Blum:2018mom,Giusti:2018mdh} are carried out with the aid of an estimate of effective models, for instance, the chiral perturbation theory (ChPT)~\cite{Aubin:2015rzx,Golterman:2017njs,Bijnens:2017esv} or the Gounaris-Sakurai (GS) parametrization~\cite{DellaMorte:2017dyu,Meyer:2018til}, to correct the FV effect on ($\simlt 6$ fm$)^3$ boxes at a long distance. In Ref.~\cite{Blum:2018mom} the leading-order ChPT estimate is added to the lattice result on a (5.4 fm$)^3$ box at the physical pion mass taking higher-order contributions of $\mathcal O(p^4)$ as a systematic error. Reference~\cite{Borsanyi:2017zdw} employs a similar strategy to add the ChPT estimate to the lattice results on (6.1--6.6 fm$)^3$ lattices around the physical pion mass but takes the systematic error conservatively. In Ref.~\cite{DellaMorte:2017dyu} the GS parametrization is used to fit the LQCD result of the vector correlator on a roughly (4 fm$)^3$ lattice at the unphysical pion mass ($m_\pi\ge 185$ MeV) with the time-slice cut of $1.1<t_{\rm cut}<1.4$ fm. References~\cite{Chakraborty:2016mwy,Giusti:2018mdh} take account of only the two-pion contributions based on the analytic estimate with ChPT. 

As pointed out in our previous study~\cite{Izubuchi:2018tdd}, it is essentially important to assess the FV effect in the LQCD calculation of $a_\mu^{\rm hvp}$ by employing the direct comparison between different volumes at the physical pion mass without any reliance on the effective models. We have made a direct evaluation of the systematic uncertainty of the FV effect using different volumes of (8.1 fm$)^3$ and (5.4 fm$)^3$ near the physical pion mass ($m_\pi=$135--145 MeV). The difference between the results on two volumes was found to be larger than the ChPT estimate, though the statistical error was so large that they are consistent within 1 $\sigma$ error bar. In this article, we perform a more precise comparison with ChPT using a lattice larger than (10 fm)$^4$ at the physical pion mass, which are a subset of the PACS10 configurations~\cite{Ishikawa:2018jee} generated by the PACS Collaboration. We also investigate the lattice cutoff effect by comparing the results at two different cutoffs of $a^{-1}= 2.33$ and 3.09 GeV keeping the physical volume larger than (10 fm)$^4$. We finally estimate an extrapolated value of $a_\mu^{\rm hvp}$ in the continuum limit and compare it with other recent LQCD results.

This paper is organized as follows. In Sec.~\ref{sec:background} we explain the notation and the LQCD methodology to calculate $a_\mu^{\rm hvp}$. Lattice parameters and numerical method are explained in Sec.~\ref{sec:setup}. The results for the FV effect and the lattice cutoff effect are presented in Secs.~\ref{sec:FV} and \ref{sec:latart}, respectively. In Sec.~\ref{seq:mom_rep} we numerically check the consistency between the results with the coordinate and momentum integration schemes. In Sec.~\ref{sec:discussion} we discuss our result in comparison with the phenomenological estimate and other recent LQCD results.  The conclusion and an outlook are summarized in Sec.~\ref{sec:summary}.

\section{Methodology}\label{sec:background}

\subsection{Momentum-space integration scheme}
\label{subsec:momentum-integration}
$a_\mu^{\rm hvp}$ is given by the integral of the vacuum polarization function (VPF) $\Pi(Q^2)$ from zero to infinity in terms of the spacelike momentum squared $Q_{\rm M}^2=-Q^2<0$:
\begin{eqnarray}
  &&a_\mu^{\rm hvp} = \Big(\frac{\alpha_e}{\pi}\Big)^2\int^\infty_0 dQ^2 K_E(Q^2)\hat\Pi(Q),
  \label{eq:a_mu_infinite}\\
  &&\hat\Pi(Q) \equiv \Pi(Q)-\Pi(0),
  \label{eq:Pihat}\\
  &&K_E(s) = \frac{1}{m_\mu^2}\hat sZ^3(\hat s)\frac{1-\hat s Z(\hat s)}{1+\hat sZ^2(\hat s)},\\
  && Z(\hat s) = -\frac{\hat s-\sqrt{\hat s^2+4\hat s}}{2\hat s},\quad \hat s = \frac{s}{m_\mu^2},
\end{eqnarray}
where $\hat\Pi(Q)$ is scheme independent due to a subtraction of scheme-dependent $\Pi(0)$. The QED kernel $K_E(s)$, which is obtained by the one-loop perturbation with $\alpha_e=1/137.03599914$ and $m_\mu=105.6583745$ MeV~\cite{Lautrup:1971jf}, has a sharp peak at $Q^2\approx (\sqrt{5}-2)m_\mu^2 = 0.003$ GeV$^2$ and a rapid falloff for $Q^2\rightarrow 0$. The vacuum polarization function can be extracted from a factorization of the vacuum polarization tensor $\Pi_{\mu\nu}(Q)$, which is given by the Fourier transformation of the vector-vector current correlator,
\begin{eqnarray}  
  \Pi_{\mu\nu}(Q) &\equiv& \sum_{x}e^{iQx}\langle V_\mu^{\Gamma}(x)V_\nu^{\Gamma'}(0)\rangle= (Q^2\delta_{\mu\nu}-Q_\mu Q_\nu)\Pi(Q),
  \label{eq:ftt_vvt}
\end{eqnarray}
where the index $\Gamma$ in the superposition of the vector current $V_\mu$ denotes two choices of lattice operators. One is the local current with ${\Gamma'}={\rm L}$:
\begin{equation}
  V_\mu^{\rm L}(x) = Z_V\bar q(x)\gamma_\mu q(x) \label{eq:v_loc}
\end{equation}
with $Z_V$ being the renormalization constant, and the other is the conserved current with $\Gamma={\rm C}$:
\begin{eqnarray}
  V_\mu^{\rm C}(x) = \frac{1}{2}\Big[ \bar q(x+a\hat\mu)(1+\gamma_\mu)U_\mu^{\dag}(x)q(x) - \bar q(x)(1-\gamma_\mu)U_\mu(x)q(x+a\hat\mu)\Big]
\end{eqnarray}
in the point-splitting form with the link variable $U_\mu(x)$, which preserves the lattice Ward-Takahashi identity  $\sum_\mu\nabla_\mu^*V_\mu=0$ with the backward differential $\nabla^*(x,y)=\delta_{x,y}-\delta_{x-\hat\mu,y}$ in the naive Wilson quark action. Note that the lattice local vector current and conserved current are not $\mathcal O(a)$ improved in this study. In Sec.~\ref{sec:latart}, we investigate the scaling violation for both currents.

The expression of $\Pi_{\mu\nu}(Q)$ in Eq.~(\ref{eq:ftt_vvt}) has extra contributions of $\mathcal O((aQ)^n)$ with $n\ge 2$ due to the Lorentz symmetry breaking on the discretized space-time in LQCD. After subtracting these lattice artifacts~\cite{Shintani:2008ga,Shintani:2010ph,Hudspith:2018bpz} $\hat\Pi(Q^2)$ computed with LQCD is consistent with the perturbative representation of the Adler function~\cite{Baikov:2008jh} in high $Q^2>1$ GeV$^2$ except for the nonperturbative objects such as the $d$-dimensional operator condensate term given by $\langle O_d\rangle/Q^{2d}$ appearing in the operator product expansion (OPE)~\cite{Braaten:1991qm}. For the actual computation of $a_\mu^{\rm hvp}$, the LQCD evaluation of the integral of Eq.~(\ref{eq:a_mu_infinite}) can be replaced by the perturbative one in the high $Q^2$ region from  some particular point of $Q^2_{\rm pQCD}$ to infinity. Practically, the integrand for $Q^2_{\rm pQCD}>1$ GeV$^2$ in Eq.~(\ref{eq:a_mu_infinite})  gives minor contribution to the total $a_\mu^{\rm hvp}$ so that the OPE contribution should be negligible. We will discuss it later.

In LQCD we need to evaluate $\Pi(0)$ by the extrapolation of VPF to the zero-momentum limit. Since the minimum momentum in LQCD is defined as $Q_{\rm min}=2\pi/L$, a large volume allows us to perform the qualified extrapolation with less uncertainty of fitting procedures.  Once $\Pi(0)$ is determined, the momentum integral of Eq.~(\ref{eq:a_mu_infinite}) is straightforwardly performed with the extrapolation function in the low-energy regime, and we can add the perturbative QCD formula in high-energy regime of $Q^2> Q_{\rm pQCD}$. As pointed out above, the choice of $Q_{\rm pQCD}^2>$1 GeV$^2$ gives only a minor contribution to the total $a_\mu^{\rm hvp}$.

In our analysis the $Q^2$ integral of Eq.~(\ref{eq:a_mu_infinite}) is split into the fit region,  the lattice data region, and the perturbative QCD (pQCD) region:
\begin{eqnarray}
  [a_\mu^{\rm hvp}]_{\rm Mom} &=& \int^{Q^2_{\rm fit}}_0 dQ^2 W_q(Q^2)\hat\Pi_f(Q)\nonumber\\
  &&+ \frac 1 2\sum_{Q^2_n=Q^2_{\rm fit}}^{Q^2_n<Q^2_{\rm pQCD}}\Big(W_q(Q_{n+1}^2)\hat\Pi_{\rm lat}(Q_{n+1}) + W_q(Q_n^2)\hat\Pi_{\rm lat}(Q_n)\Big)(Q^2_{n+1}-Q^2_n)\nonumber\\
  &&+ \int^\infty_{Q^2_{\rm pQCD}} dQ^2 W_q(Q^2)\hat\Pi_{\rm pQCD}(Q),
  \label{eq:a_mu_mom}\\
  W_q(s) &\equiv& \Big(\frac{\alpha_e}{\pi}\Big)^2 K_E(s),\\
  \hat\Pi_{\rm lat}(Q) &=& \Pi_{\rm lat}(Q) - \Pi(0),
\end{eqnarray}
where we define the lattice momentum $Q_n=2\pi n/L$ with integer $n=\{0,1,...,L/a-1\}$ and $\hat\Pi_{\rm pQCD}$ is an analytic form in pQCD. $\hat\Pi_f(s)\equiv \Pi_f(s)-\Pi(0)$ is a functional form with the fitting ansatz, which is used for the extrapolation of the lattice data to obtain $\Pi(0)$. We utilize three types of fitting functions,
\begin{eqnarray}
  &\textrm{(Pad\'e ansatz)}&\Pi_f^{\rm Pade[1,1]}(s) = \Pi(0) + \frac{s X_0}{s+X_1},\\
  && \Pi_f^{\rm Pade[2,1]}(s) = \Pi(0) + \frac{s X_0}{s+X_1}+sX_2,\\
  &\textrm{(Linear approx.)}&\Pi_f^{\rm linear}(s) = \Pi(0) + s Y 
\end{eqnarray}
with the fitting parameters $\Pi(0)$, $X_0$, $X_1$ $X_2$. Note that we use the same form of Pad\'e approximation as in Refs.~\cite{Aubin:2012me,DellaMorte:2017dyu}.

\subsection{Coordinate-space integration scheme}
As an alternative approach we consider the vector-vector current correlator in the coordinate space:
\begin{equation}
  C(x) = \sum_\mu\langle V_\mu(x)V_\mu(0)\rangle,
  \label{eq:vvx}
\end{equation}
where the summation over the same component of sink and source vector currents is taken. With the use of $C(x)$, $a_\mu^{\rm hvp}$ can be expressed as follows:~\footnote{See the Appendix~\ref{appendix:dev} for the derivation.}
\begin{eqnarray}
  a_\mu^{\rm hvp} &=& \Big(\frac{\alpha_e}{\pi}\Big)^2\int d^4x C(x)\int^\infty_0 d\omega^2 K_E(\omega^2)\frac{4\pi^2}{3\omega^2}\Big[ e^{iQx} - 1\nonumber\\
    && -\frac{\omega^2}{2}\lim_{\varepsilon^2=0}\Big\{\sum_{\mu\nu}(-x_\mu x_\nu)\frac{\varepsilon^2}{P_\mu P_\nu}\Big|_{P_\mu\ne0,P_\nu\ne0,\varepsilon=|P|}\Big\}\Big]_{\omega=|Q|}.
  \label{eq:a_mu_coor_gen}
\end{eqnarray}
Here there is a degree of freedom for a choice of four components in the momentum $Q_\mu$ satisfying $Q^2=\omega^2$. If we take one component is non-zero and others are zero, {\it i.e.} $Q_\mu=\{Q_\rho=|Q|,Q_{\mu\ne\rho}=0\},\,\rho=\{x,y,z,t\}$, Eq.~(\ref{eq:a_mu_coor_gen}) is simplified as
\begin{eqnarray}
  a_\mu^{\rm hvp} = \Big(\frac{\alpha_e}{\pi}\Big)^2\int dx_\rho \bar C_\rho(x_\rho)\int^\infty_0 d\omega^2 K_E(\omega^2)\frac{4\pi^2}{\omega^2}\Big[ e^{i\omega x_\rho}-1+\frac{\omega^2x_\rho^2}{2}\Big],
  \label{eq:a_mu_coor}
\end{eqnarray}
where we define
\begin{equation}
  \bar C_\rho(x_\rho) = \sum_{\mu\ne\rho}\int \Big(\prod_{\sigma\ne\rho}d{x_\sigma}\Big)  \langle V_\mu(x)V_\mu(0)\rangle.
  \label{eq:corr_x}
\end{equation}
This is a formula called time-momentum representation (TMR)~\cite{Bernecker:2011gh}, in particular, with the choice of the time direction for $x_\rho$~\footnote{We will not call ``TMR'' for Eq.~(\ref{eq:corr_x}), alternatively saying ``coordinate-space representation'' since the word ``Time'' may be a confusion.}. It is consistent with the Lorentz-covariant coordinate-space representation~\cite{Meyer:2017hjv} if the coordinate-space integral in Eq.~(\ref{eq:a_mu_coor_gen}) is transformed into the spherical and radial integrals.

In LQCD we perform a discretized coordinate-space summation of the correlator on the finite lattice volume defined as 
\begin{eqnarray}
  [a_\mu^{\rm hvp}]_{\rm lat}(r_{\rm cut}) &=& \frac1 2\sum_{r/a=0}^{r_{\rm cut}/a-1} \Big[C^{\Gamma\Gamma'}(r)W_r(r) + C^{\Gamma\Gamma'}(r+a)W_r(r+a)\Big],\label{eq:a_mu_coor_lat}\\
  W_r(r) &=& 8\alpha_e^2\int^\infty_0\frac{d\omega}{\omega}K_E(\omega^2)\Big[\omega^2r^2-4\sin^2(\omega r/2)\Big]\label{eq:wr_lat}
\end{eqnarray}
with
\begin{equation}
  C^{\Gamma\Gamma'}(x) = \sum_\mu\langle V^{\Gamma}_\mu(x)V^{\Gamma'}_\mu(0)\rangle,
  \label{eq:vvx_lat}
\end{equation}
where $r$, which denotes a distance from the source point, is regarded as the generalized expression of $x_\rho$ in Eq.~(\ref{eq:a_mu_coor}) and thus $C^{\Gamma\Gamma'}(r)$ represents $\bar C_\rho(x_\rho)$ on the lattice. This procedure introduces the systematic uncertainties due not only to the discretized summation but also to the truncation at some finite distance $r_{\rm cut}$
~\footnote{On the lattice, since the momentum is discretized, the momentum integral in Eq.~(\ref{eq:wr_lat}) should be replaced as the summation of lattice momentum squared of $\omega^2=\sum_\mu (2\pi n_\mu/L_\mu)^2$ for $n_\mu=[0,L_\mu-1]$, while we naively use the momentum integral as the continuum. This assumption may also introduce additional systematic uncertainty but taking the continuum limit and infinite volume limit, it will not be a concern.}. 

As we will explain below, the lattice used in this study is symmetric and its spatial/temporal extension is large enough to control the finite volume effect and the backward propagation state (BPS) effect investigated in Ref.~\cite{Izubuchi:2018tdd}.  We can perform the integral of Eq.~(\ref{eq:a_mu_coor}) [summation of Eq.~(\ref{eq:a_mu_coor_lat})] for each direction of $\rho=x,y,z,t$, which allows us to increase the statistics by four times without much computational cost. 

\section{Calculation details}\label{sec:setup}

\subsection{Configurations}
\label{subsec:configurations}

We use two subsets of the PACS10 configurations, which are generated with the stout-smeared $\mathcal O(a)$-improved Wilson-clover quark action and Iwasaki gauge action \cite{Iwasaki:2011np} on 128$^4$ and $160^4$ lattices (spatial extension $L$ and temporal extension $T$ are symmetric) at $\beta=1.82$ and 2.00, respectively. In addition we also employ the gauge field configurations on a $64^4$ lattice at $\beta=1.82$, which are copied in the temporal direction extending $T/a$ to 128 in the FV study. Lattice parameters for these configuration sets are summarized in Table~\ref{tab:lattice}. We investigate the FV effect using the $128^4$ and $64^4$ lattices at the same lattice spacing, and the study of the cutoff effect uses the $128^4$ and $160^4$ lattices with the fixed physical volume.

The detailed description of the configuration generation on the $128^4$ and $64^4$ lattices was already given in Ref.~\cite{Ishikawa:2018jee}. Here we explain the configuration generation on the $160^4$ lattice at $\beta=2.00$. We employ the stout smearing parameter $\rho=0.1$, and the number of the smearing steps is six, which are the same as in the case of the $128^4$ lattice at $\beta=1.82$~\cite{Ishikawa:2018jee}.The improvement coefficient of $c_{\rm SW}=1.02$ is nonperturbatively determined by the Schr{\"o}dinger functional (SF) scheme following Ref.~\cite{Taniguchi:2012kk}. The hopping parameters for the light (degenerate up-down) and strange quarks ($\kappa_{\rm ud}$,$\kappa_{\rm s}$)=(0.125814,0.124925) are carefully adjusted to yield the physical pion and kaon masses ($m_\pi$,$m_K$)=(135.0 MeV,497.6 MeV) with the use of the cutoff of $a^{-1}=3.09$ GeV ($a=0.064$ fm)~\cite{PACS_ScaleSet} determined from the $\Xi$ mass $m_\Xi=1.3148$ GeV. The hopping parameter for the charm (only valence quark) is set to $\kappa_c=0.110428$ on $128^4$ lattice, and $\kappa_c=0.11452$ on $160^4$ lattice adjusted to physical point.

The degenerate up-down (ud) quarks are simulated with the domain-decomposed HMC (DDHMC) algorithm~\cite{Luscher:2003vf,Luscher:2005rx} on the $160^4$ lattice. The ud quark determinant is separated into the UV and IR parts after the even-odd preconditioning. We also apply the twofold mass preconditioning~\cite{Hasenbusch:2001ne,Hasenbusch:2002ai} to the IR part by splitting it into $\tilde F_{\rm IR}$, $F_{\rm IR}^{\prime}$ and $F_{\rm IR}^{\prime\prime}$. This decomposition is controlled by two additional hopping parameters:
$\kappa^\prime_{\rm ud}=\rho_1\kappa_{\rm ud}$ with $\rho_1=0.9997$ and 
$\kappa^{\prime\prime}_{\rm ud}=\rho_1\rho_2\kappa_{\rm ud}$ with $\rho_2=0.9940$. ${\tilde F}_{\rm IR}$ is derived from the action preconditioned with
$\kappa^\prime_{\rm ud}$. The ratio of two preconditioners with 
$\kappa^\prime_{\rm ud}$ and $\kappa^{\prime\prime}_{\rm ud}$ gives
$F^{\prime}_{\rm IR}$.  
$F^{\prime\prime}_{\rm IR}$ is from the heaviest preconditioner 
with $\kappa^{\prime\prime}_{\rm ud}$. 
In the end the force terms  consist of  the gauge force $F_{\rm g}$, the UV force $F_{\rm UV}$ and the three IR forces $F_{\rm IR}^{\prime\prime}$, $F_{\rm IR}^{\prime}$ and $\tilde F_{\rm IR}$.  The IR forces are obtained with the mixed precision nested BiCGStab method for the quark solver~\cite{Aoki:2008sm}.
We adopt the multiple time scale integration scheme~\cite{Sexton:1992nu}  in the molecular dynamics (MD) steps. 
The associated step sizes are controlled by a set of integers $(N_0,N_1,N_2,N_3,N_4)$: $\delta\tau_{\rm g}=\tau/N_0 N_1 N_2 N_3 N_4,$ 
$\delta\tau_{\rm UV}=\tau/N_1 N_2 N_3 N_4,$ 
$\delta\tau^{\prime\prime}_{\rm IR}=\tau/N_2 N_3 N_4,$ 
$\delta\tau^{\prime}_{\rm IR}=\tau/N_3 N_4,$ 
$\delta{\tilde \tau}_{\rm IR}=\tau/N_4$ with $\tau=1.0$.
Our choice of $(N_0,N_1,N_2,N_3,N_4)=(8,2,2,2,20)$ for the $160^4$ lattices results in 82\% acceptance rates. 
The strange quark is simulated with the RHMC algorithm~\cite{Clark:2006fx} choosing the force approximation range of [min,max]=[0.000190,1.90] with $N_{\rm RHMC}=10$ and $\delta\tau_{\rm s}=\delta\tau^{\prime\prime}_{\rm IR}$ for the step size. 

The renormalization constant $Z_V$ for the local vector current operator in Eq.~(\ref{eq:v_loc}) depends on the lattice cutoff scale. We obtain  $Z_V=0.95153(76)$ at $a^{-1}=2.33$ GeV ($\beta=1.82$) with the SF scheme~\cite{Ishikawa:2015fzw}, and $Z_V=0.9673(19)$ at $a^{-1}=3.09$ GeV ($\beta=2.00$) from the nucleon form factor. Note that we observe a good consistency between the results of $Z_V$ determined by the SF scheme and the nucleon form factor~\cite{Shintani:2018ozy}. The physical observables are measured at every 10 trajectories on $128^4$ and $64^4$, and every 5 trajectories on $160^4$. The statistical error is estimated by the jackknife analysis with 1, 4, and 5 jackknife binsizes for the $128^4$, $160^4$, and $64^4$ lattices respectively~\cite{Ishikawa:2018jee}.

\begin{table}
  \begin{center}
    \caption{Summary of the lattice parameters for the gauge field configurations used in this work.  ($^*$)Four rotational degrees of freedom is taken into account. Originally ten configurations are generated. ($^{**}$)$64^4$ gauge configurations are copied in the temporal direction extending $T/a$ to 128.}\label{tab:lattice}
    \begin{tabular}{ccccccc}
      \hline\hline      
      & Refs. & $L/a$ $[L]$& $T/a$ $[T]$& $a^{-1}$ (GeV) & $m_\pi$ (MeV) & \#configs\\
      \hline
      \multirow{2}{*}{PACS10} &\cite{Ishikawa:2018jee} & 128 [10.8 fm]& 128 [10.8 fm]& 2.333(18) & 135 & 21\\
      &  & 160 [10.3 fm]& 160 [10.3 fm] & 3.087(30) & 135 & 40$^*$\\
      \hline
      & \cite{Ishikawa:2018jee,Izubuchi:2018tdd} & 64 [5.4 fm] & 64 [5.4 fm] & 2.333(18) & 139 & 187\\
      & \cite{Izubuchi:2018tdd} & 64 [5.4 fm] & 128$^{**}$ [10.8 fm] & 2.333(18) & 139 & 86\\
      \hline\hline
    \end{tabular}
  \end{center}
\end{table}
 
\subsection{AMA with deflated SAP preconditioning}
\label{subsec:ama}
     
The precision of the light flavor vector-vector current correlator in the infrared (IR) regime, which is the small $Q^2$ region in Eq.~(\ref{eq:a_mu_infinite}) or the long distance region from the source location in Eq.~(\ref{eq:a_mu_coor_lat}), has a vital importance  to achieve a less than 1\% level of accuracy for $a_\mu^{\rm hvp}$ with LQCD. As in the previous study~\cite{Izubuchi:2018tdd} we utilize the optimized all-mode-averaging (AMA) technique~\cite{Blum:2012uh,Shintani:2014vja,vonHippel:2016wid} to make an efficient calculation of the vector-vector current correlator in LQCD. For the AMA approximation~\cite{vonHippel:2016wid,Izubuchi:2018tdd}, we use the parameter set illustrated in Table~\ref{tab:ama_params}. As shown in Refs.~\cite{vonHippel:2016wid,Izubuchi:2018tdd}, the combination of AMA with the deflated Schwartz alternative procedure (SAP) preconditioning~\cite{Luscher:2007se} achieves the remarkable performance on the large lattice, and it then allows us a precise calculation of $a_\mu^{\rm hvp}$, especially, in a long distance region. In fact the condition number in the AMA method with the deflated SAP preconditioning does not have large volume dependence~\cite{Izubuchi:2018tdd} since the low mode is effectively eliminated by the projection with a deflation filed so that the computational cost to solve the light quark propagator does not increase even if the lattice size is enlarged. Although the computational cost of generation of deflation fields is increased in large lattice size, it takes only a few percent of the total computational cost~\cite{Izubuchi:2018tdd} in deflated SAP preconditioning. This provides us with another advantage to avoid consuming large storage space to save the low-lying mode.

\begin{table}
  \begin{center}
    \caption{The parameter of AMA approximation on $64^4$, $128^4$, and $160^4$ lattices. ``SAP domain'' column denotes the size of SAP domain, and ``Deflation'' column denotes the number of deflation fields on the deflated SAP preconditioning. ``Iteration'' denotes the stopping iteration of General Conjugate Residual (GCR) method.}\label{tab:ama_params}
    \begin{tabular}{ccccc}
      \hline\hline
      Quark & Lattice & SAP domain & Deflation & Iteration \\
      \hline
      \multirow{3}{*}{Light} &
      $64^4$ & $4^4$ & 30 & 5 \\
      & $128^4$ & $8^4$ & 50 & 7 \\
      & $160^4$ & $10^4$ & 50 & 7 \\
      \hline
      \multirow{3}{*}{Strange} &
      $128^4$ & $8^4$ & 46 & 5 \\
      & $160^4$ & $10^4$ & 30 & 5 \\
      \hline\hline
    \end{tabular}
  \end{center}
\end{table}

\begin{figure}
  \begin{center}
    \includegraphics[width=70mm]{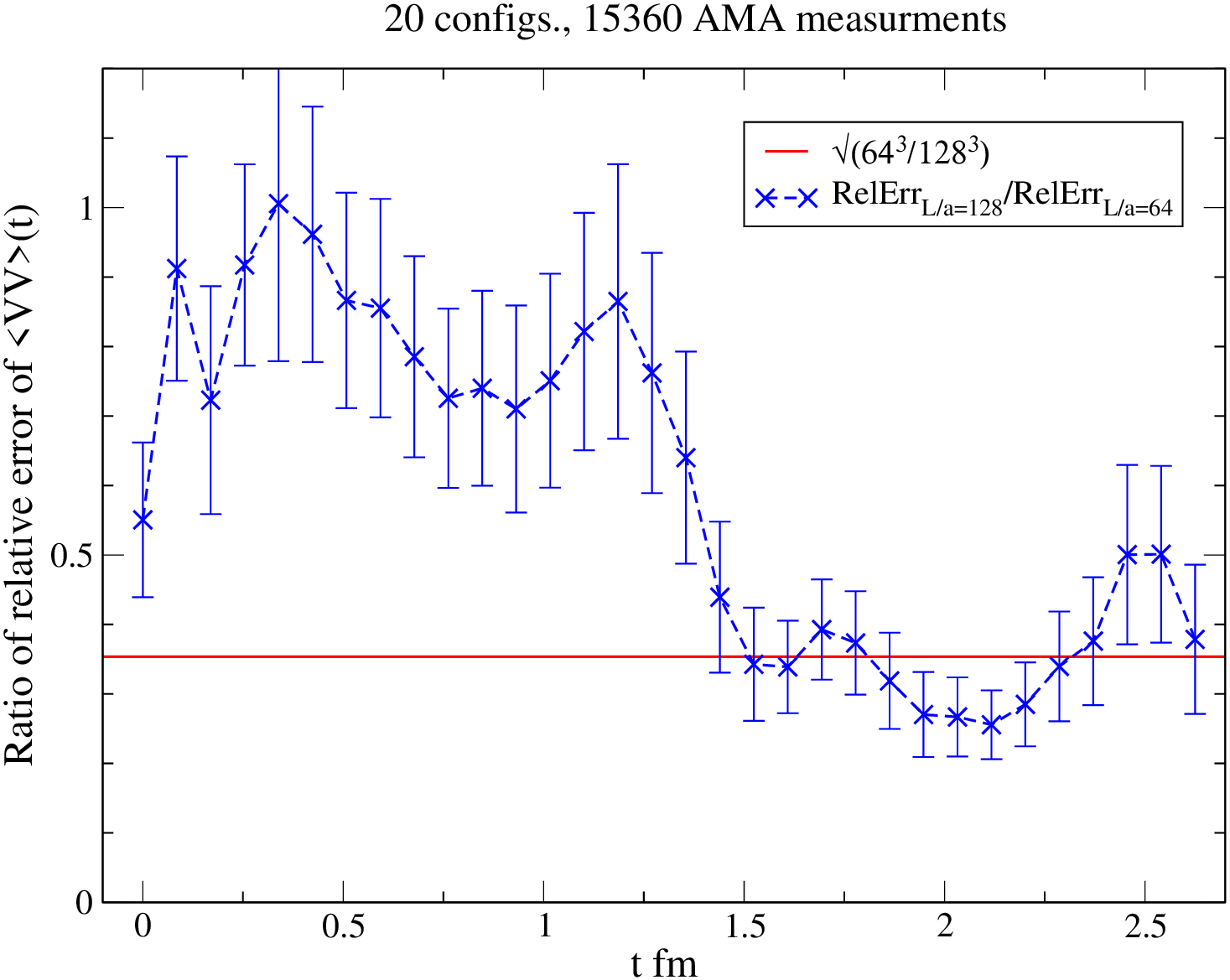}
    \hspace{3mm}
    \includegraphics[width=71mm]{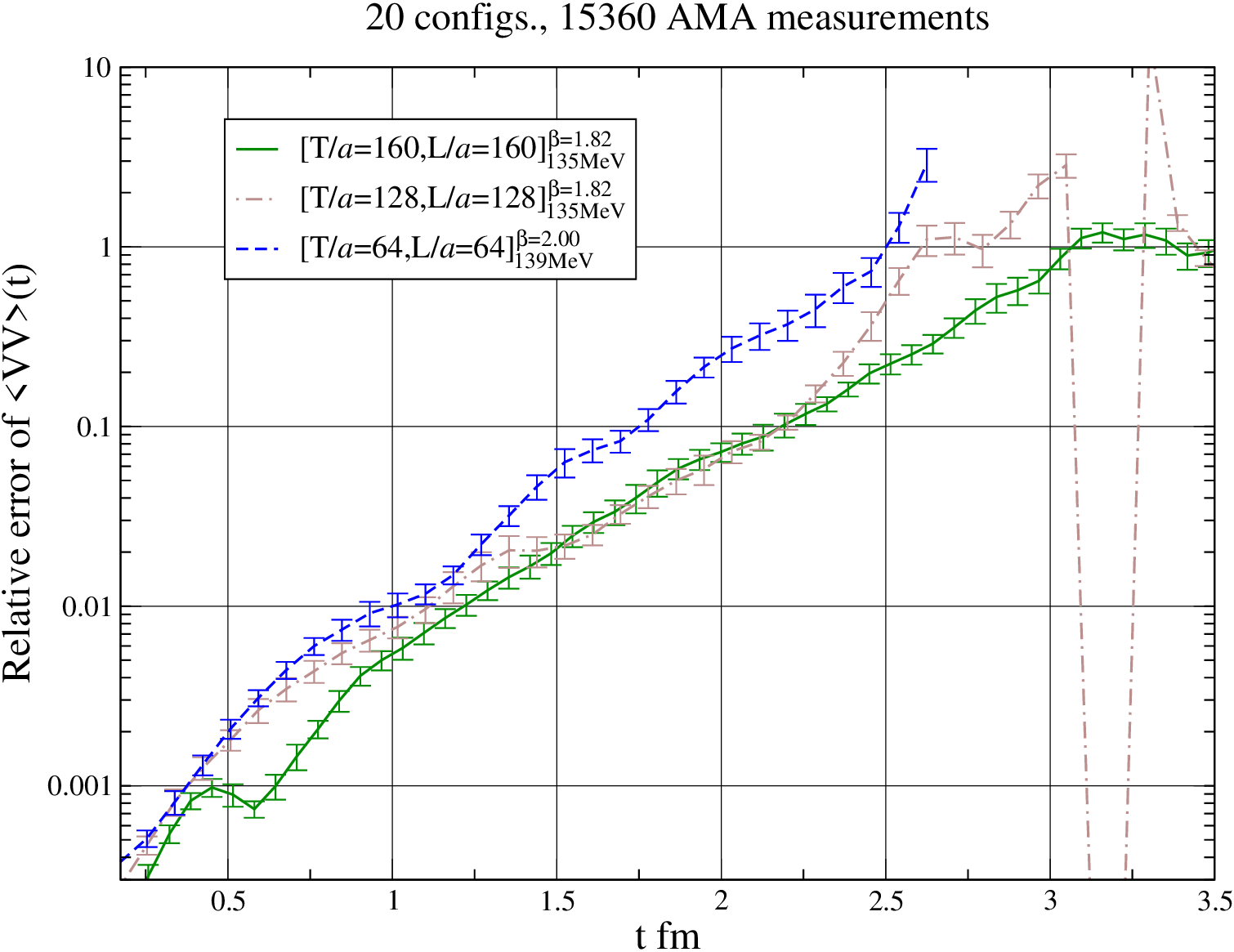}
    \caption{(Left) Ratio of relative error for the vector-vector current correlator $C^{\rm CL}(t)$ on 128$^4$ and 64$^4$ lattices using the same number of measurements. The straight line shows the expected volume scaling. (Right) Relative error of the vector-vector current correlator on 128$^4$ and 64$^4$ lattices in $a^{-1}=2.33$ GeV and 160$^4$ lattice at $a^{-1}=3.06$ GeV.}\label{fig:relerr}
  \end{center}
\end{figure}

In the left panel of Fig.~\ref{fig:relerr} we show the volume scaling for the relative error of the correlator at the physical pion mass. This is more robust test of volume scaling than the previous study~\cite{Izubuchi:2018tdd}, where there might be possible contamination due to the pion mass difference between two volumes. From this plot, one can see that the ratio of the relative error between $64^4$ and $128^4$ lattices has a consistent behavior with the expected scaling value of $\sqrt{64^3/128^3}$ in a long distance region $t\simgt 1.5$ fm, which means that the use of a large volume can significantly reduce the statistical error, especially for the IR regime. As illustrated in the right panel of Fig.~\ref{fig:relerr}, we also observe the universal behavior for the relative error of the vector-vector current correlator at different cutoff scales on the same physical volume. This feature is also  expected from the volume scaling hypothesis for the statistical error.

\subsection{Multihadron state contributions}
\label{subsec:multihadron}

\begin{figure}
  \begin{center}
    \includegraphics[width=100mm]{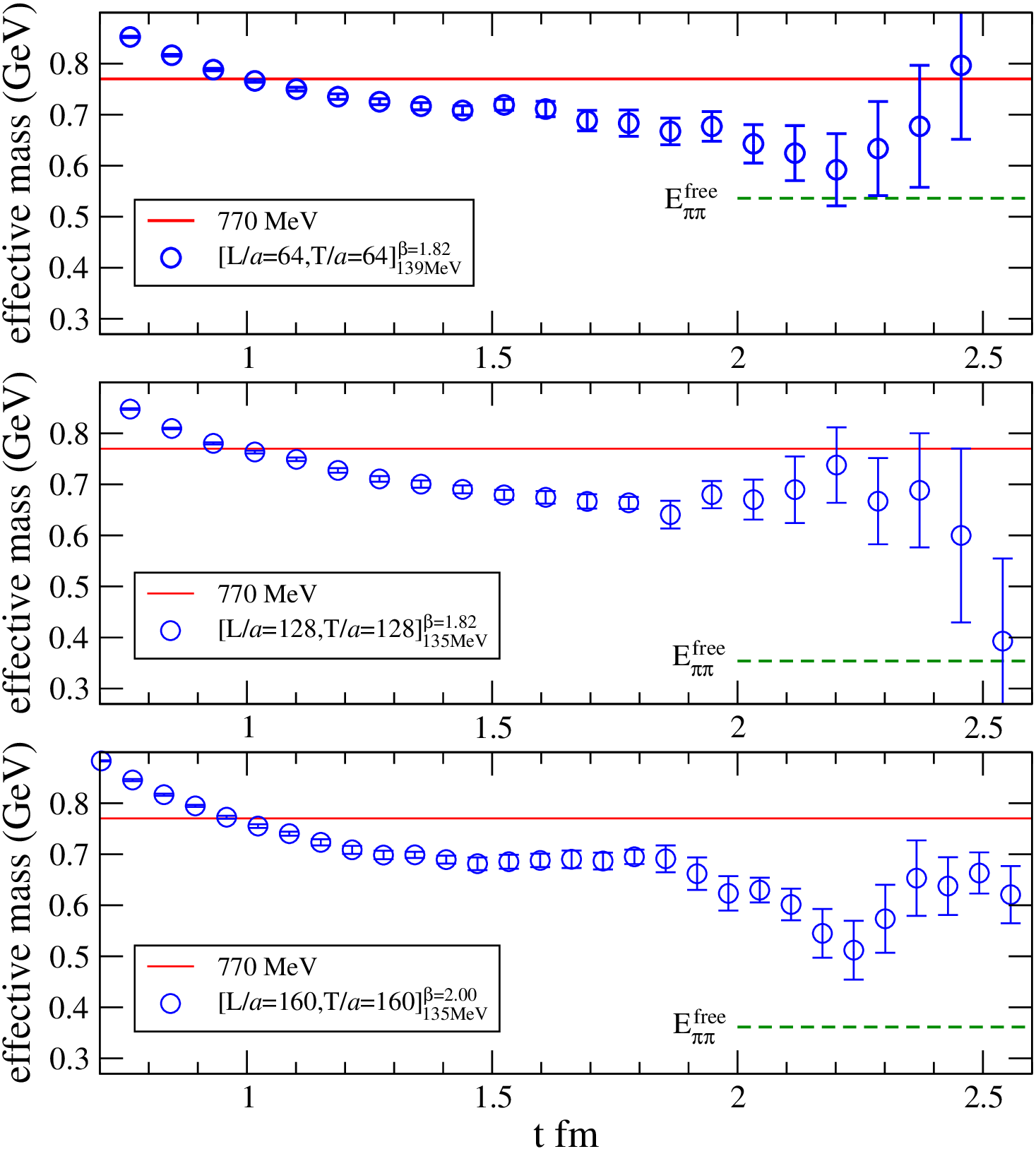}
    \caption{Effective mass for the vector-vector current correlator at the physical pion mass on 64$^4$ (top), 128$^4$ (middle) and 160$^4$ (bottom) lattices. Solid lines denote the physical $\rho$ meson mass and dashed ones are for the free two-pion energies $E_{\pi\pi}^{\rm free}=2\sqrt{m_\pi^2+(2\pi/L)^2}$ on each lattice volume.}\label{fig:mass}
  \end{center}
\end{figure}

Using our large lattice ensembles at the physical pion mass, the multihadron state contributions, mostly the two-pion state, are correctly involved in the vector-vector current correlator. Figure~\ref{fig:mass} plots the effective mass of the vector-vector current correlator with $(\Gamma,\Gamma')=$(L,C) on each ensemble. The $\rho$ meson is allowed to decay into the energetic pions on those ensembles, since the two-pion state energy $2\sqrt{m_\pi^2+(2\pi/L)^2}$ is much lower than the $\rho$ meson mass $m_\rho=770$ MeV. We observe that the effective mass goes down below 770 MeV around $t\approx 1$ fm and stays above the energy level of $2\sqrt{m_\pi^2+(2\pi/L)^2}$ in the large $t$ region on each ensemble. This is a clear indication of the existence of lower energy state than the $\rho$ meson mass in the region of $t\simgt 1$ fm, which is dominated by the multihadron state contributions. 

\section{Numerical results}

With the use of the gauge field configurations explained in Sec.~\ref{sec:setup}, we perform a systematic study of uncertainties stemming from the FV effect, the cutoff effect and the integration scheme dependence in LQCD calculation of $a_\mu^{\rm hvp}$. For the FV effect we directly compare the results for the coordinate-space integral of Eq.~(\ref{eq:a_mu_coor_lat}) obtained on the $L/a=128$ and $L/a=64$ lattices at the same cutoff scale of $a^{-1}=2.33$ GeV. The cutoff effect is investigated by calculating the coordinate-space integral of Eq.~(\ref{eq:a_mu_coor_lat}) on the $128^4$ and $160^4$ lattices keeping the physical lattice volume constant. We also discuss the operator dependence of the cutoff effect for the local and conserved vector currents. Finally we examine the consistency between the coordinate- and momentum-space integration schemes on the $128^4$ ((10.8 fm$)^4$) lattice at the physical pion mass.

\subsection{Finite volume effect}\label{sec:FV}

\begin{figure}
  \begin{center}
    \includegraphics[width=110mm]{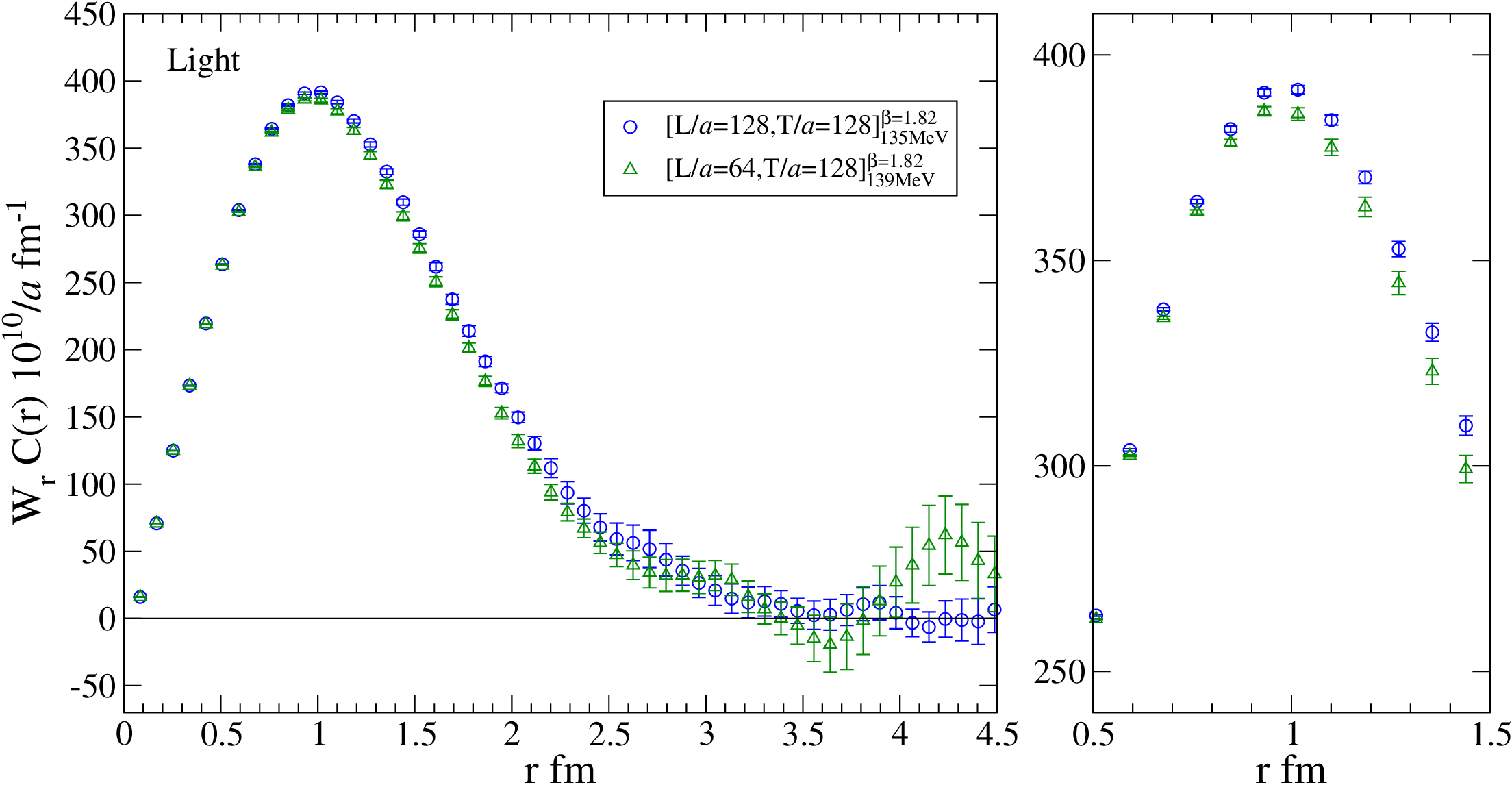}
    \caption{Comparison of $C^{\rm LL}(r)W_r(r)$ in Eq.~(\ref{eq:a_mu_coor_lat}) between different spatial volumes with $L/a=128$ and 64 in light quark sector. }\label{fig:diff_amu_voldep_128t}
  \end{center}
\end{figure}

\begin{figure}
  \begin{center}
    \includegraphics[width=75mm]{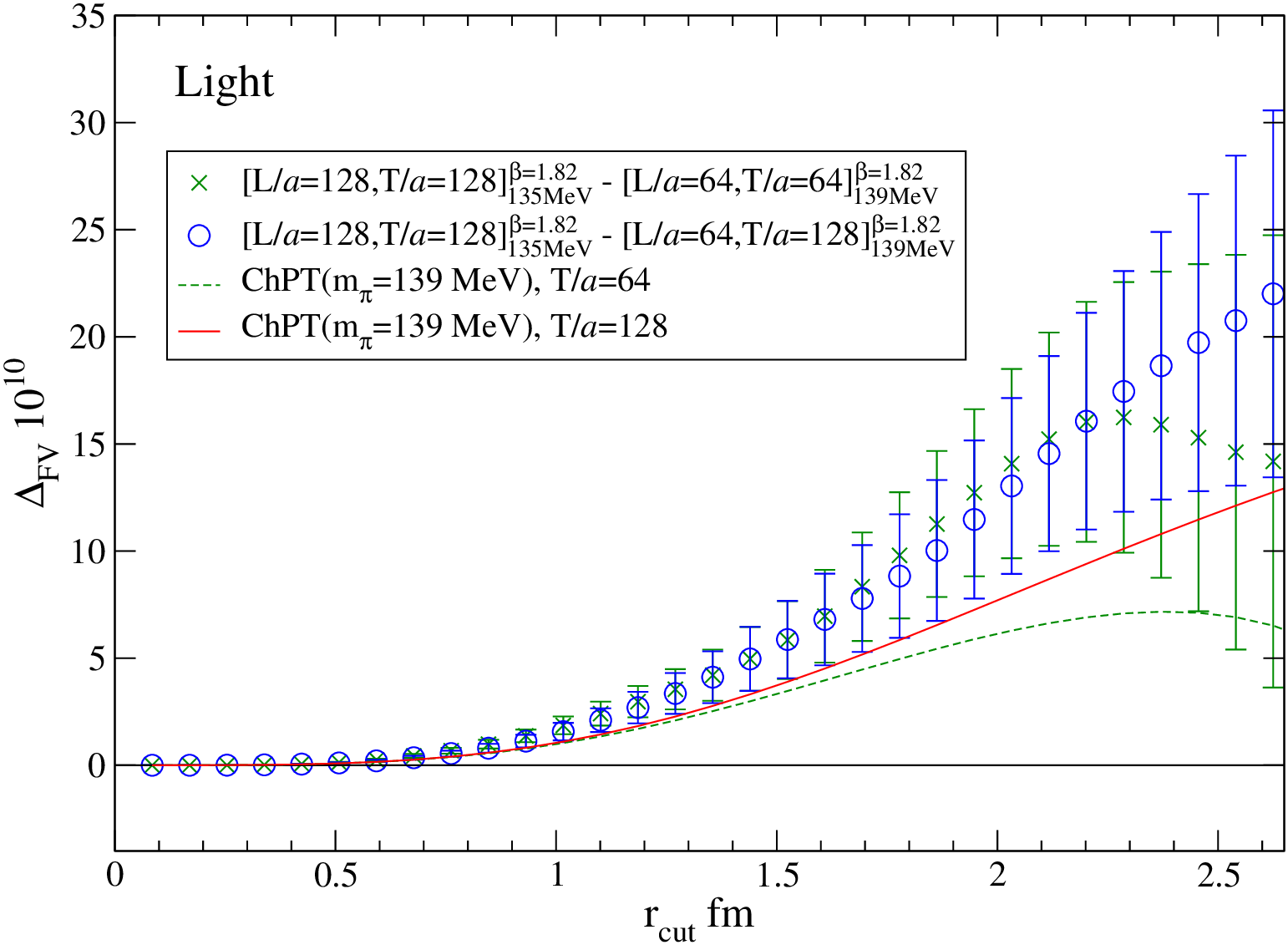}
    \hspace{3mm}
    \includegraphics[width=75mm]{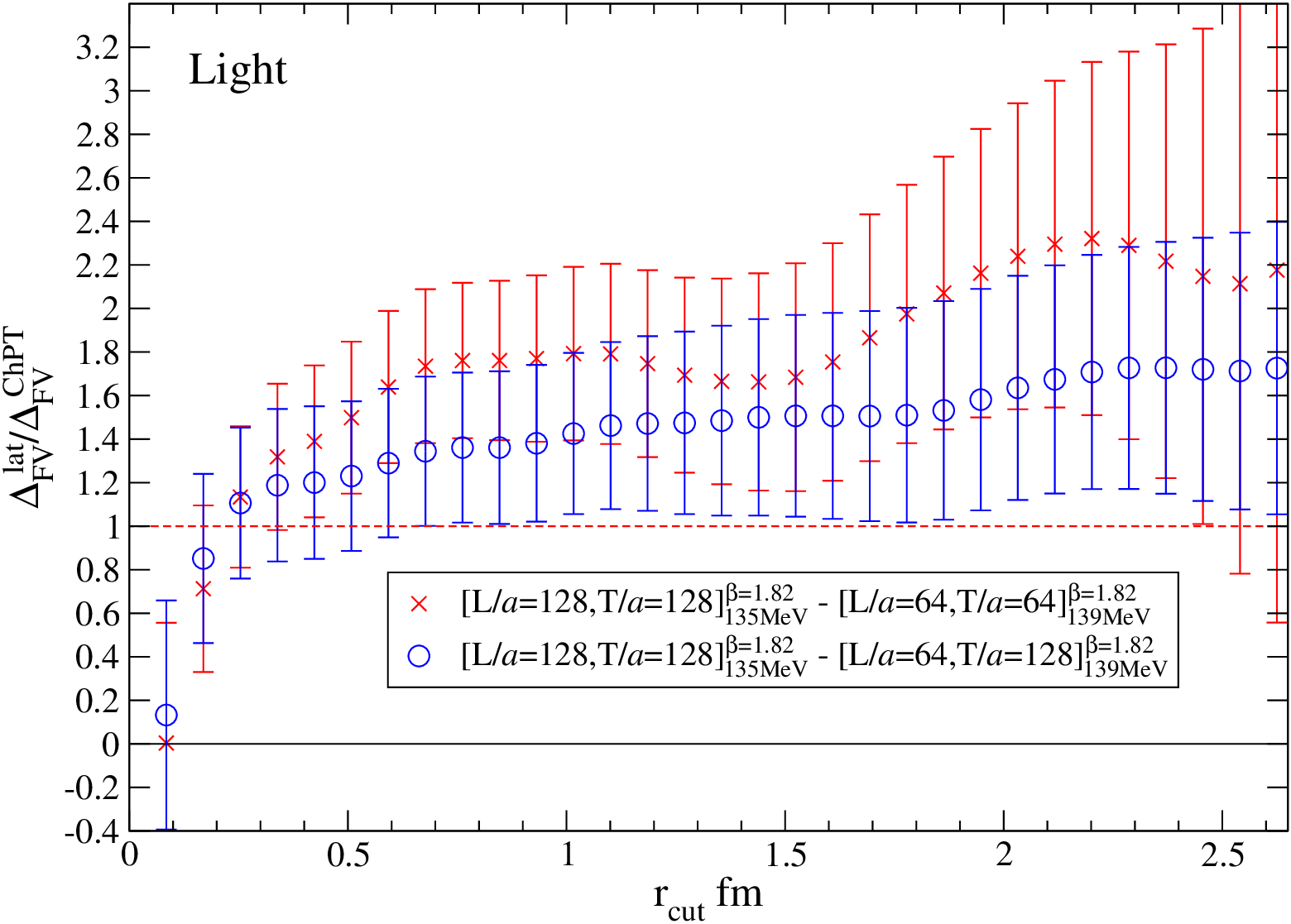}
    \caption{(Left) Difference of $a_\mu^{\rm hvp}(r_{\rm cut})$ on the 128$^4$, $64^4$, and 64$^3\times 128$ lattices in the light quark sector. The hopping parameters are the same on both lattices. The solid (dashed) curve denotes the leading order of the ChPT prediction for the FV effect between (10.8 fm$)^3$ and (5.4 fm$)^3$ spatial volumes with $m_\pi= 135$ MeV on $[L/a=128,T/a=128]$ lattice and 139 MeV on $[L/a=64,T/a=128]$ ($[L/a=64,T/a=64]$) lattice. (Right) Ratio of the FV effect between the LQCD and ChPT estimates with the same symbol as left panel.}\label{fig:ratio_amu_voldep}
  \end{center}
\end{figure}

Figure~\ref{fig:diff_amu_voldep_128t} shows the comparison of integrand in Eq.~(\ref{eq:a_mu_coor_lat}) between $L/a=128$ and $L/a=64$ lattices. For the latter section, we extend $T/a$ to 128 by copying the $64^4$ lattice in the temporal direction so that we can eliminate the BPS wrapping around temporal direction observed in our previous study~\cite{Izubuchi:2018tdd} and discussed below. We remark that, although the $64^4$ lattice configurations are generated at the same hopping parameter as for the $128^4$ lattice, the measured pion mass $m_\pi=139$ MeV on the $64^4$ lattice is slightly heavier than $m_\pi=135$ MeV on the $128^4$ lattice due to the FV effect~\cite{Ishikawa:2018jee}. In the right panel of Fig.~\ref{fig:diff_amu_voldep_128t}, one can see that the integrand has a clear tendency in which the magnitude for integrand increases when $L/a$ is enlarged from 64 to 128. 
The left panel of Fig.~\ref{fig:ratio_amu_voldep} plots the FV effect defined as
\begin{eqnarray}
  \Delta_{\rm FV} = \big[a_\mu^{\rm hvp}(r_{\rm cut})\big]^l_{L/a=128}-\big[a_\mu^{\rm hvp}(r_{\rm cut})\big]^l_{L/a=64},
\end{eqnarray}
which shows that the magnitude is larger than the leading order ChPT having the same sign with the ChPT prediction~\cite{Aubin:2015rzx}. In this figure we also make a comparison using the result on $64^4$ lattice. One can see that in IR regime, more than $r_{\rm cut}=2.3$ fm, the BPS effect may be involved into FV correction as enlarged $a_\mu^{\rm hvp}$ effect on 64$^4$, and it then turns out to be additional systematic uncertainty. Use of extended temporal direction as $64^3\times 128$ thus plays an important role to avoid such a BPS effect from FV correction. In order to clarify the discrepancy  we plot the ratio of the FV effect between LQCD and ChPT at each $r_{\rm cut}$ in the right panel of Fig.~\ref{fig:ratio_amu_voldep}. One can observe that the LQCD data tends to become larger than the ChPT prediction from $r\approx 1$ fm, and this tendency does not change even if $r_{\rm cut}$ increases, though the statistical error becomes larger. 

The discrepancy of FV effect between LQCD and ChPT in the light quark sector is estimated as
\begin{equation}
  \Delta_{\rm FV}^{\rm lat}/\Delta_{\rm FV}^{\rm ChPT} = \left\{\begin{array}{cc}
    2.16(66) & [\textrm{at $r_{\rm cut}\simeq 2.0$ fm on $64^4$ lattice}],\\        
    1.74(71) & [\textrm{at $r_{\rm cut}\simeq 2.6$ fm on 64$^3\times$128 lattice}],\\
  \end{array}\right.
  \label{eq:delta_FV}
\end{equation}
on $L=5.4$ fm at the physical pion mass. Comparing $T/a=64$ and 128, as one can also see in Fig.~\ref{fig:ratio_amu_voldep}, even at $r\simeq 2$ fm, there is a significant contribution of BPS regarded as an additional FV effect. Our result in Eq.~(\ref{eq:delta_FV}) indicates that the actual FV effect tends to be larger than the ChPT prediction, which may provide useful information on other recent LQCD results using ChPT or another analysis to correct the FV effect for $a_\mu^{\rm hvp}$ on (4 to 5 fm$)^3$ box~\cite{Chakraborty:2016mwy,Borsanyi:2017zdw,Blum:2018mom,Giusti:2018mdh}\footnote{$r_{\rm cut}\simeq 2.6$ fm is the maximum point of the window method in Ref.~\cite{Blum:2018mom}, and it then means that there still may be a large FV correction}.

\begin{figure}
  \begin{center}
    \includegraphics[width=100mm]{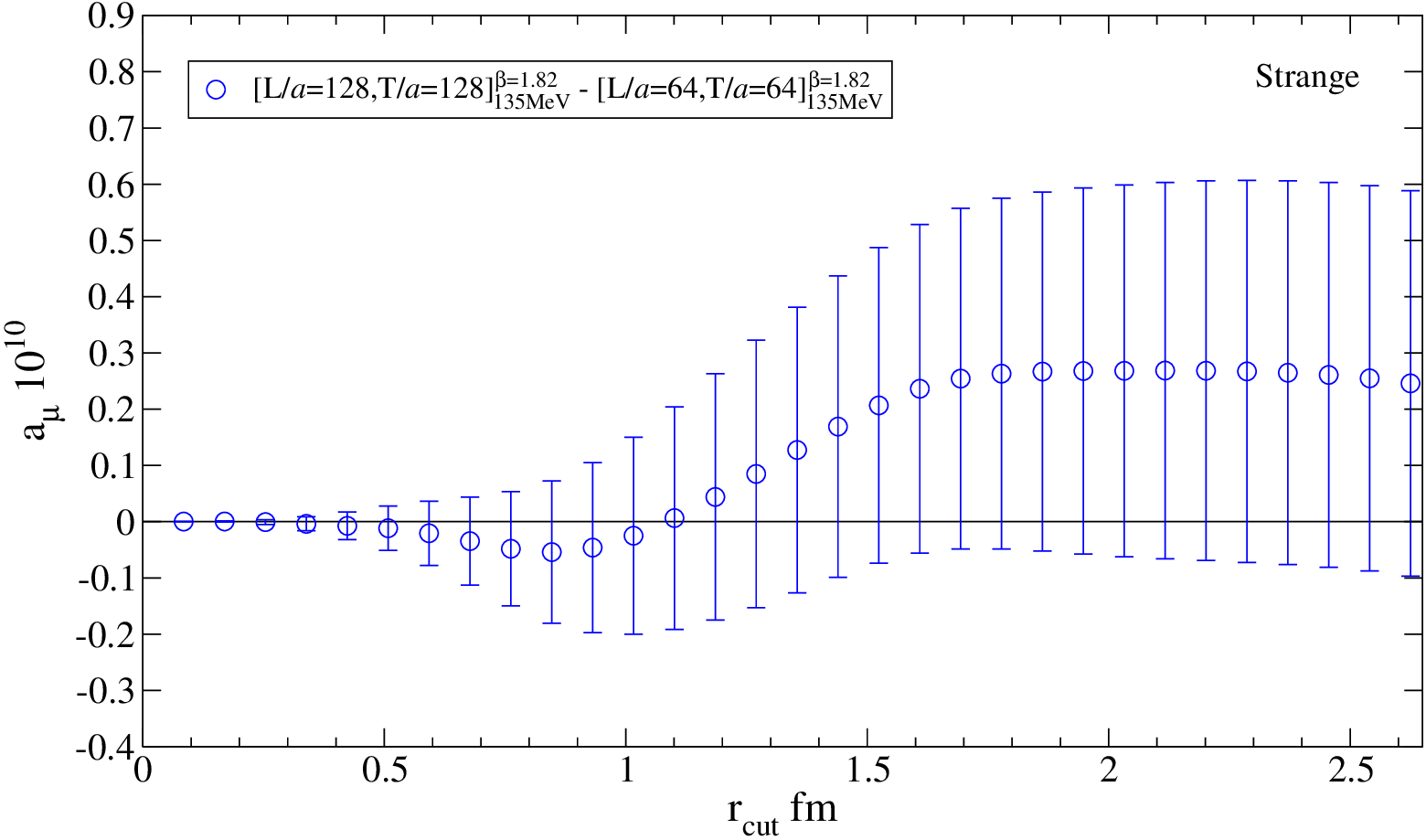}
    \caption{Difference of $a_\mu^{\rm hvp}(r_{\rm cut})$ on 128$^4$ and 64$^4$ lattices in strange quark sector.}\label{fig:diff_amu_voldep_s}
  \end{center}
\end{figure}

The similar analysis is made for the strange quark sector $[a_\mu^{\rm hvp}]^s$. Figure~\ref{fig:diff_amu_voldep_s} shows little FV effect for $[a_\mu^{\rm hvp}]^s$ as expected from the fact that the strange quark mass is much heavier than the light one. 

\subsection{Cutoff effect}\label{sec:latart}

\begin{figure}
  \begin{center}
    \includegraphics[width=120mm]{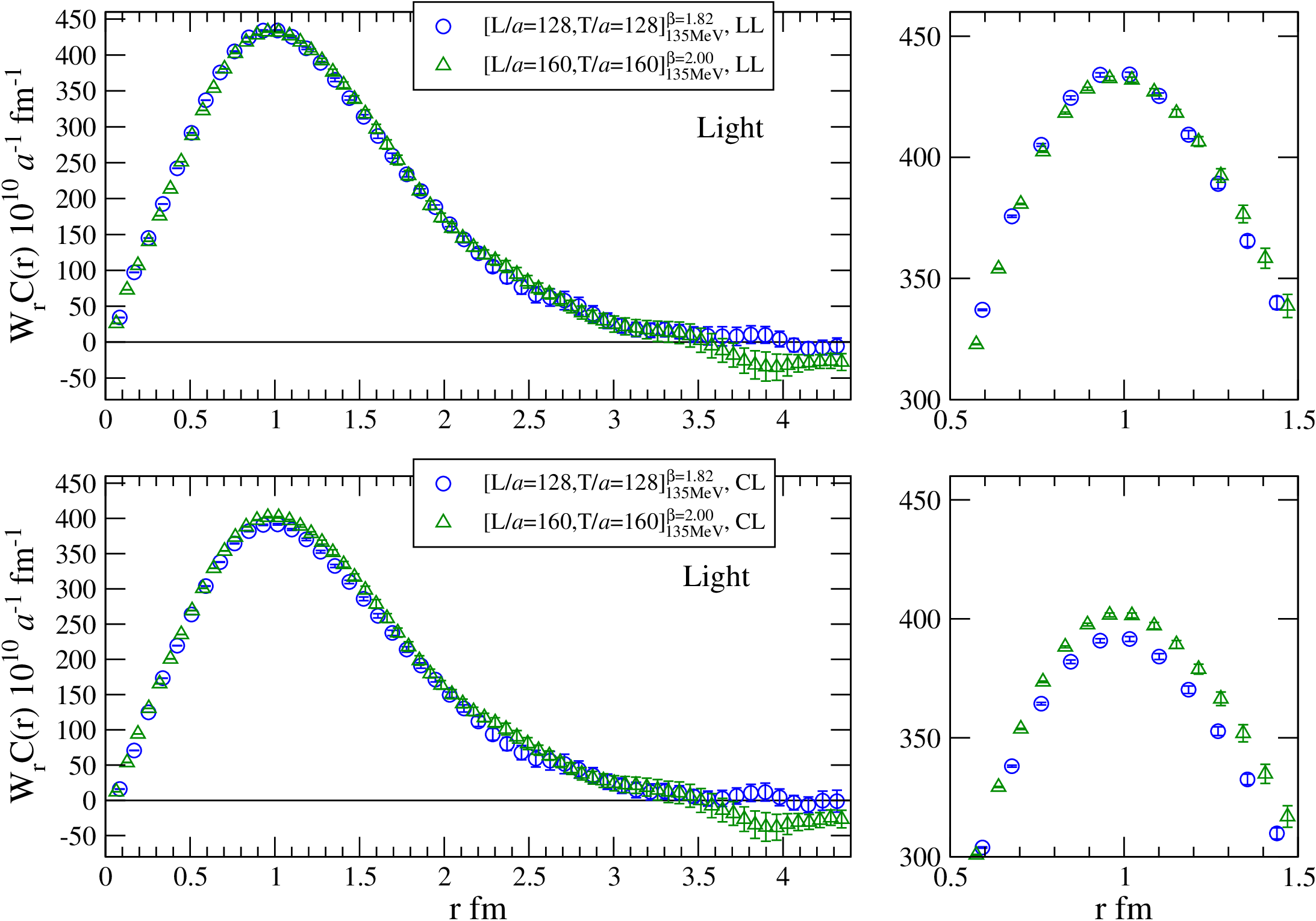}
    \caption{$C^{\rm LL}(r)W_r(r)$ (top) and $C^{\rm CL}(r)W_r(r)$ (bottom) in Eq.~(\ref{eq:a_mu_coor_lat}) in light quark sector as a function of distance $r$ on 128$^4$ lattice at $a^{-1}=2.33$ GeV  (circles) and on 160$^4$ lattice at $a^{-1}=3.06$ GeV (triangles).}\label{fig:amu_tdep_local}
  \end{center}
\end{figure}

\begin{figure}
  \begin{center}
    \includegraphics[width=70mm]{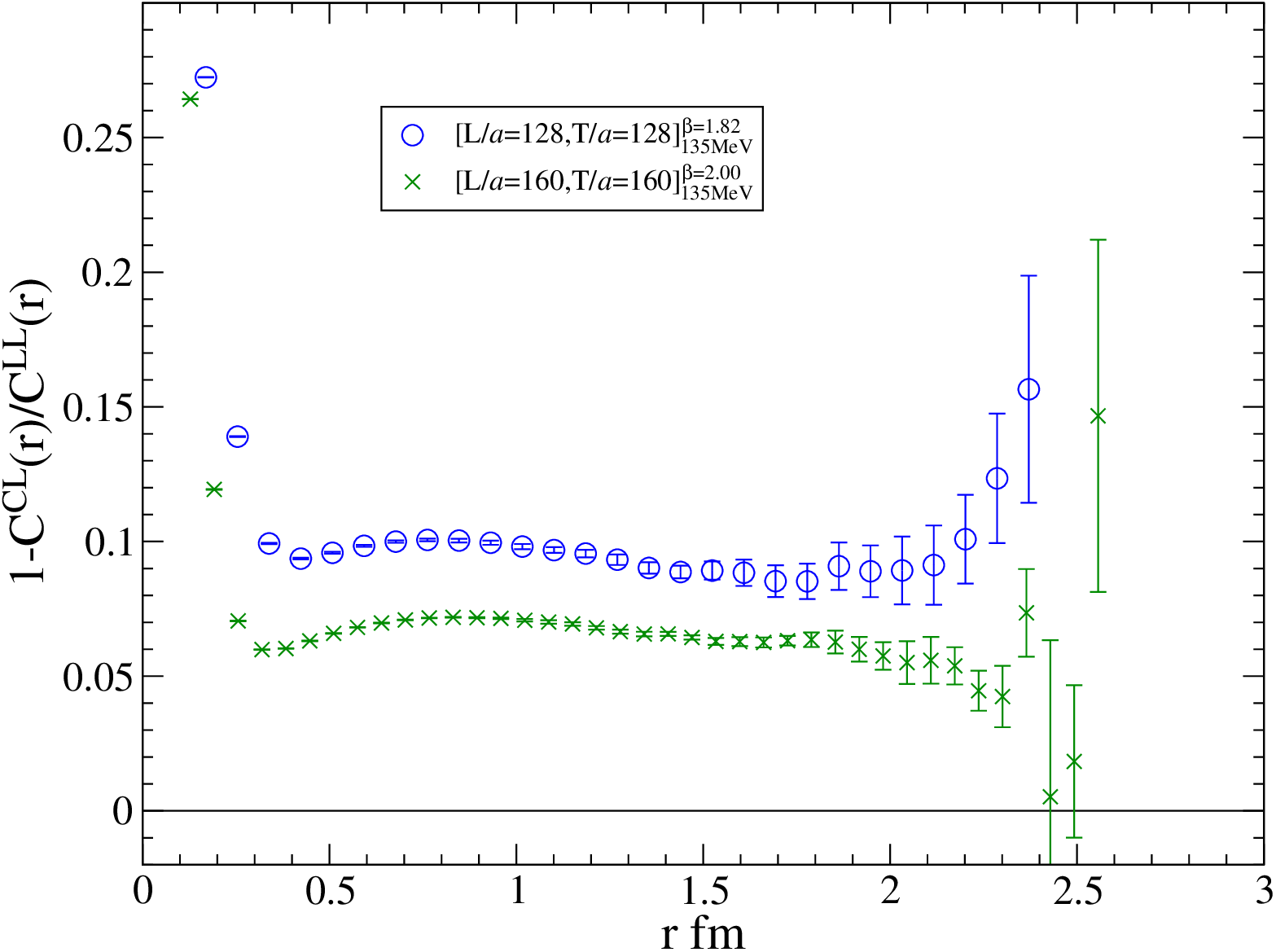}
    \hspace{3mm}
    \includegraphics[width=73mm]{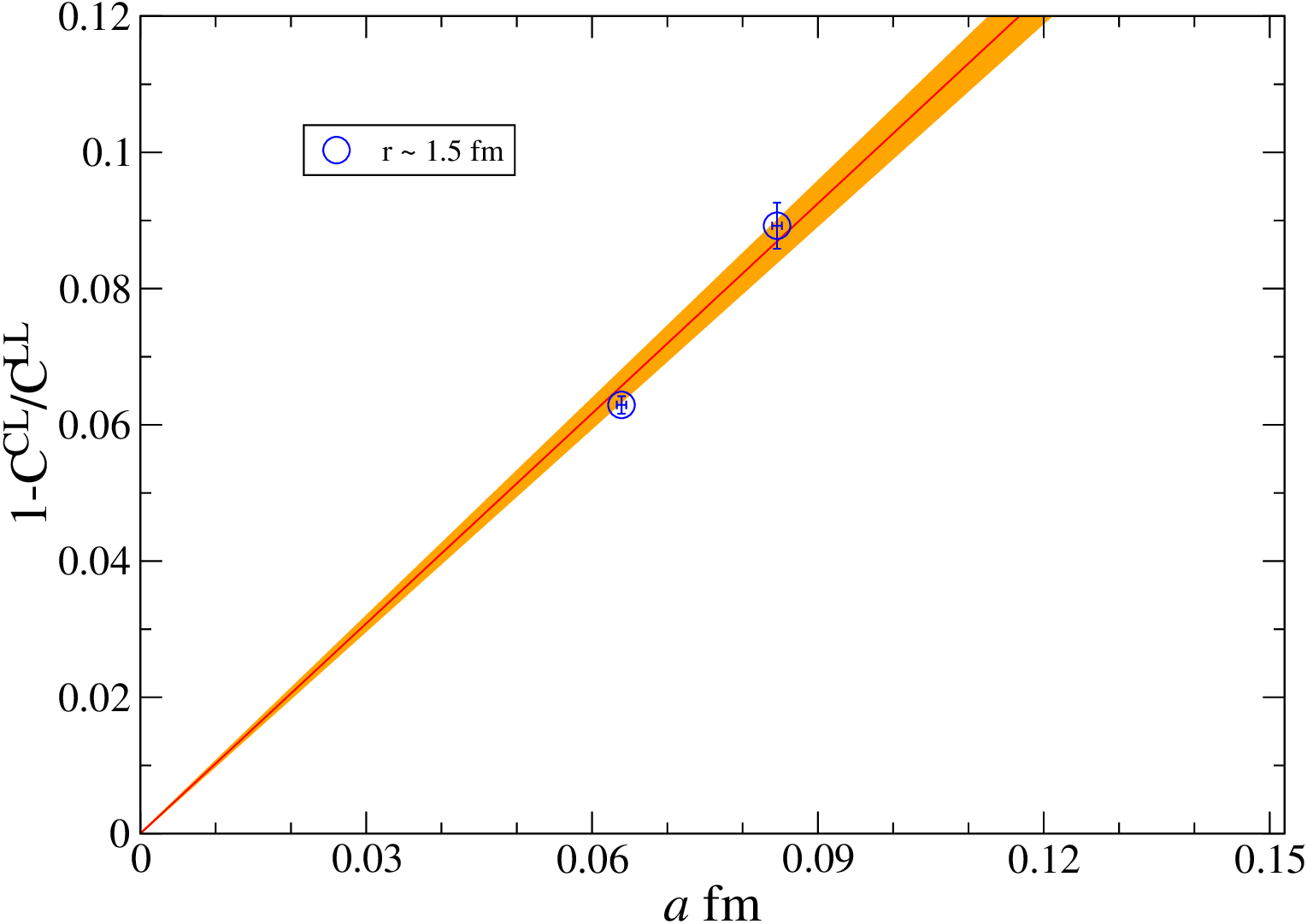}    
    \caption{(Left) $r$ dependence of $\Delta_r(r)\equiv 1-C^{\rm CL}(r)/C^{\rm LL}(r)$ on 128$^4$ lattice at $a^{-1}=2.33$ GeV (circle) and 160$^4$ lattice at 3.06 GeV (cross). (Right) The cutoff dependence of $\Delta_r(r)$ at $r \simeq 1.5$ fm. The straight line and band denote the central value and statistical error of linear fitting function of $\Delta_r(r)\propto a$.}\label{fig:amu_local_ratio}
  \end{center}
\end{figure}

In Fig.~\ref{fig:amu_tdep_local} we plot $C^{\Gamma\Gamma'}(r)W_r(r)$ in Eq.~(\ref{eq:a_mu_coor_lat}) at two different cutoff scales of $a^{-1}=2.33$ GeV and 3.09 GeV on the same physical volume over (10 fm$)^4$ at the physical pion mass. We compare the cutoff effect in two types of the vector-vector current correlators with $(\Gamma,\Gamma')=$(L,L) and (C,L) for the sink and source vector current operators in Eq.~(\ref{eq:vvx_lat}). We observe that $C^{\rm LL}(r)W_r(r)$ at different cutoff scales well agree with each other, whereas the sizable deviation is found for $C^{\rm CL}(r)W_r(r)$ from $r\simeq$0.5 fm. Our LQCD results show the $C^{\rm LL}(r)$ correlator has smaller cutoff effect than the $C^{\rm CL}(r)$ one. In order to make a quantitative measurement of the discrepancy between two types of the correlators, we plot the normalized difference defined as
\begin{equation}
  \Delta_r(r)\equiv 1-C^{\rm CL}(r)/C^{\rm LL}(r)
\end{equation}
in Fig.~\ref{fig:amu_local_ratio}. The quantity shows a clear deviation from zero, and its magnitude is reduced for the finer lattice. Around $r\approx1.5$ fm we obtain
\begin{eqnarray}
  &&\Delta_r(r\approx1.5{\rm fm})= 
  \left\{\begin{array}{cc}
  0.089(3) & (a^{-1}=2.33\,{\rm GeV})\\
  0.063(1) & (a^{-1}=3.09\,{\rm GeV})
  \end{array} \right.
\end{eqnarray}
and their ratio $\Delta_r^{1/a=3.06{\rm GeV}}/\Delta_r^{1/a=2.33{\rm GeV}}= 0.71(3)$ is comparable to the cutoff ratio of $a^{-1}_{128^4}/a^{-1}_{160^4}=[2.333(18)\,{\rm GeV}]/[3.089(30)\,{\rm GeV}]$=0.76(1) (also see the right panel of Figure~\ref{fig:amu_local_ratio}). This suggests that the LQCD result with $C^{\rm CL}(r)$ is affected by the $\mathcal O(a)$ correction due to a significant cutoff effect on the conserved (point-splitting) current. 

\begin{figure}
  \begin{center}
    \includegraphics[width=70mm]{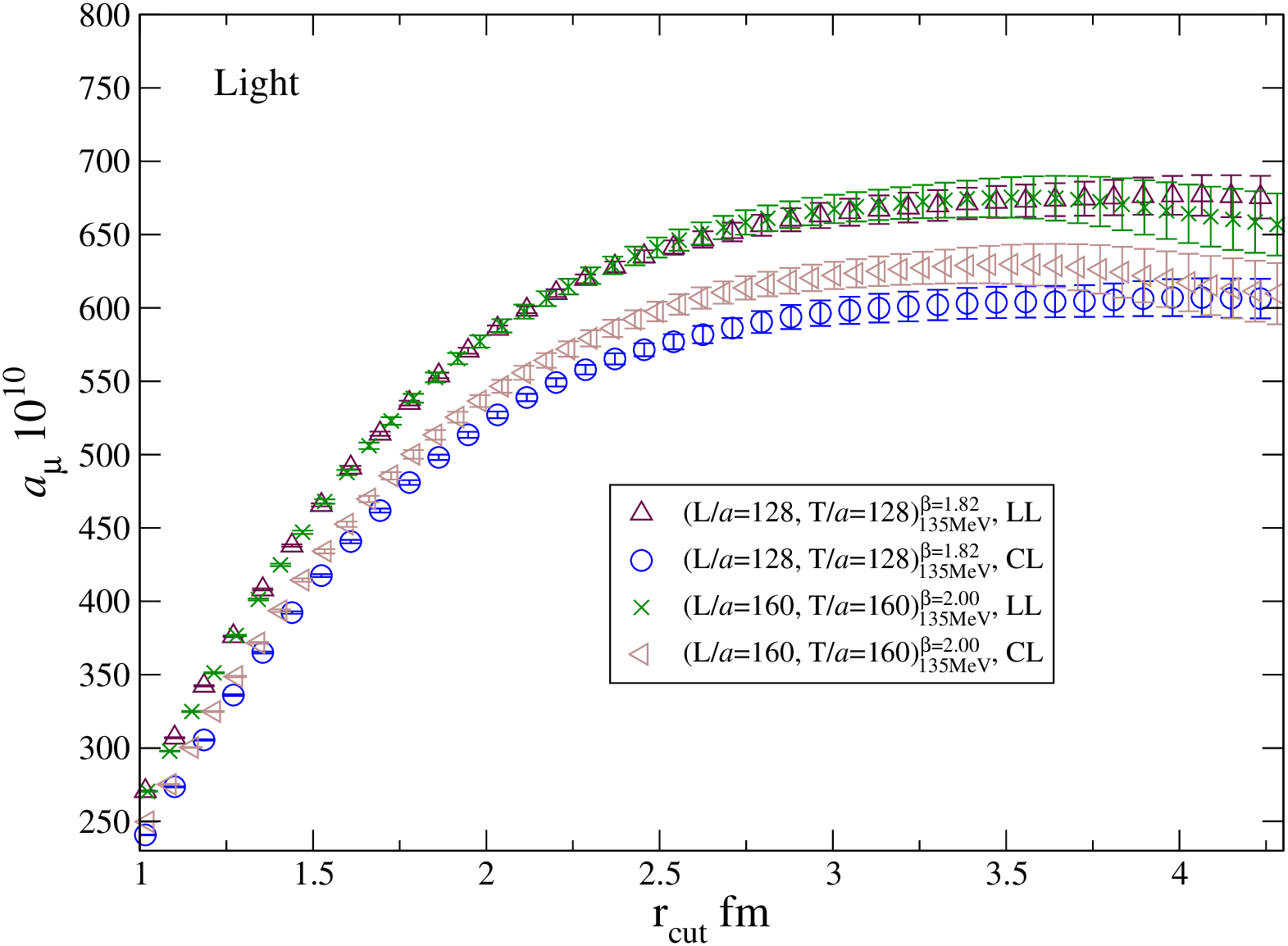}
    \hspace{3mm}
    \includegraphics[width=70mm]{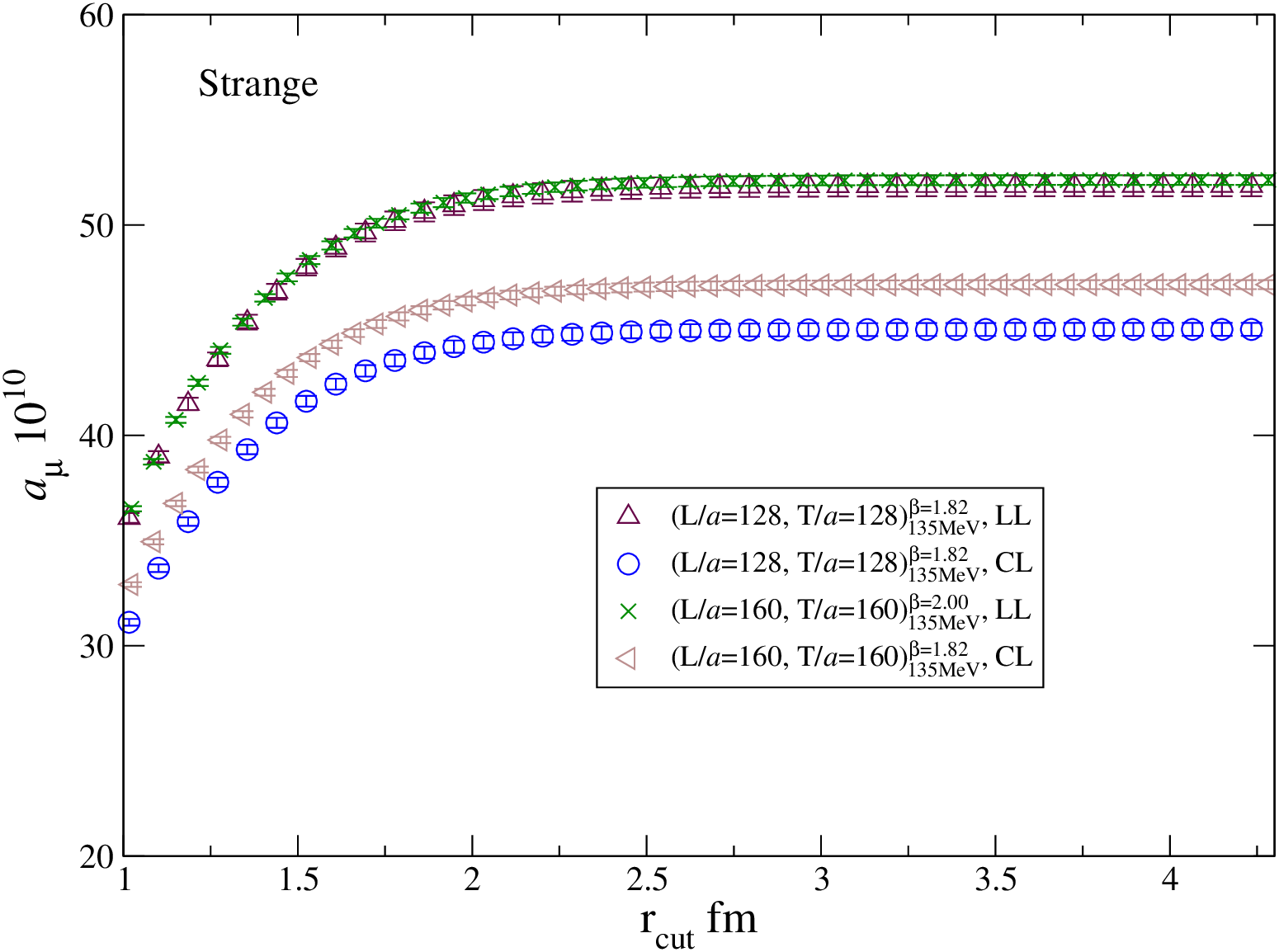}
    \caption{LQCD results for $[a_\mu^{\rm hvp}]_{\rm lat}(r_{\rm cut})$ of Eq.~(\ref{eq:a_mu_coor_lat}) in (L,L) and (C,L) channels on 128$^4$ lattice at $a^{-1}=2.33$ GeV and on 160$^4$ lattice at $a^{-1}=$3.09 GeV. Left (right) figure pertains to the light (strange) quark sector.}\label{fig:amu_local}
  \end{center}
\end{figure}

In Fig.~\ref{fig:amu_local} we plot $r_{\rm cut}$ dependence for $[a_\mu^{\rm hvp}]_{\rm lat}(r_{\rm cut})$ in the light and strange quark sectors. They asymptotically reach  constant values around $r_{\rm cut}\simgt 3.5$ fm without large statistical fluctuation. In both the light and strange quark sectors, $[a_\mu^{\rm hvp}]_{\rm lat}$ in the (L,L) channel at two cutoff scales agree within 1-$\sigma$ statistical error, while there is  10--11\% cutoff effect to affect $[a_\mu^{\rm hvp}]_{\rm lat}$ in the (C,L) channel at $a^{-1}=2.33$ GeV.

\begin{figure}
  \begin{center}
    \includegraphics[width=100mm]{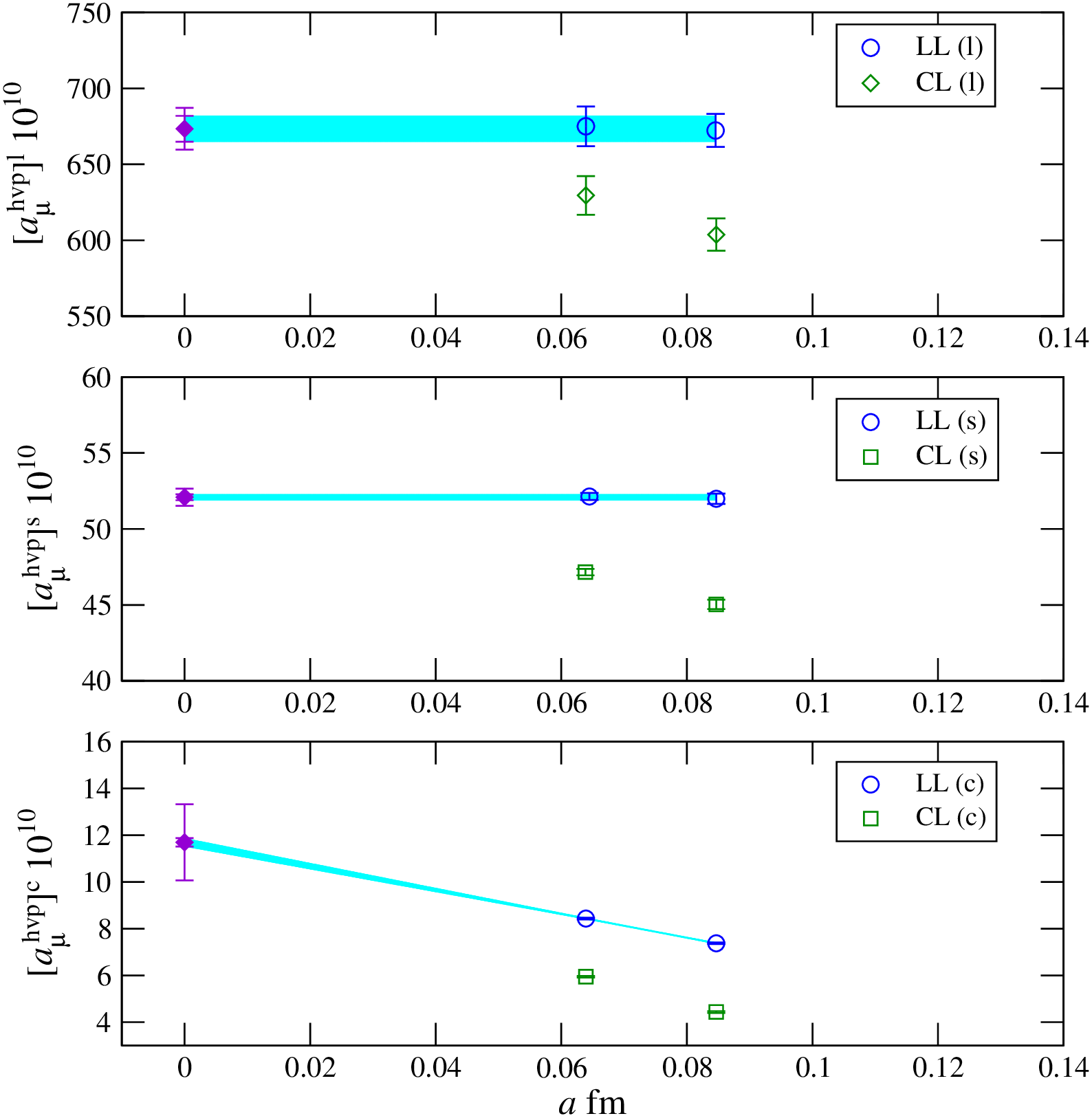}
    \caption{Cutoff dependence of $[a_\mu^{\rm hvp}]^l_{\rm lat}(r_{\rm cut})$ (top), $[a_\mu^{\rm hvp}]^s_{\rm lat}(r_{\rm cut})$ (middle) and $[a_\mu^{\rm hvp}]^c_{\rm lat}(r_{\rm cut})$ (bottom) in (L,L) and (C,L) channels with $r_{\rm cut}\approx3.5$ fm. The extrapolated result in the continuum limit (diamond) has two kinds of errors: inner one is statistical and outer one denotes the total error including the systematic error explained in the text.}\label{fig:amu_adep}
  \end{center}
\end{figure}

We summarize the scaling properties for  $[a_\mu^{\rm hvp}]^l_{\rm lat}$, $[a_\mu^{\rm hvp}]^s_{\rm lat}$ and $[a_\mu^{\rm hvp}]^c_{\rm lat}$ at two cutoff scales and their continuum extrapolations in Fig.~\ref{fig:amu_adep}, where the LQCD results at each cutoff scale are obtained by choosing $r_{\rm cut}\approx3.5$ fm. One can see that the (L,L) channel has rather small cutoff effect, which is not significant in the currently statistical precision, compared to the (C,L) channel in the light and strange quark sectors. Note that the local vector current we used here is not an $\mathcal O(a)$ improvement; however, in our lattice setup, the $\mathcal O(a)$ cutoff effect for local current is automatically suppressed, and hence such an $\mathcal O(a)$ improvement is not required. This is expected from that the magnitude of $c_A$, which is the $\mathcal O(a)$ improvement factor for local axial vector current, is almost zero when computed by the SF scheme \cite{Ishikawa:2015fzw} on the same lattice setup, and correspondingly $c_V$ will be similar order of magnitude. In addition, we check that the contribution of higher dimension operator $\bar q(\partial \sigma)_\mu q$, which is usually used for the $\mathcal O(a)$ improvement of local vector current, is an order of magnitude smaller than that of the naive local vector current. On the other hand, for conserved current without $\mathcal O(a)$ improvement, it is clear that the correction of the $\mathcal O(a)$ cutoff effect is needed to reduce the cutoff uncertainty even for our lattice setup. In our analysis, even though there are only two variations of lattice cutoff, it would be acceptable to use a constant fit of (L,L) channel for $[a_\mu^{\rm hvp}]^l_{\rm lat}$ and $[a_\mu^{\rm hvp}]^s_{\rm lat}$ to take the continuum extrapolation, and we omit (C,L) channel to avoid the additional systematic uncertainty due to fitting with $\mathcal O(a)$ and higher cutoff correction. 

The systematic error is evaluated by taking the maximum difference between the central value obtained by the constant fit and the linearly extrapolated values in the (L,L) channels with the ansatz of the $\mathcal O(a)$ term including the error of the lattice cutoff itself. The magnitude of the systematic error is comparable to that of the statistical one in the light and strange quark sectors. For the charm quark sector the bottom panel of  Fig.~\ref{fig:amu_adep} shows the large cutoff effect due to the $\mathcal O(am_c)$ contribution even in the (L,L) channel. So we take the linearly extrapolated value in the (L,L) channel as the central value in the continuum limit, and its systematic error of the $\mathcal O(a^2)$ contribution is naively estimated as $(c_1a)^2/c_0$, where $c_1$ is defined in the fit result of the linear extrapolation $c_0+c_1 a$. Further analysis of the cutoff effect in the charm sector will be done by adding the data on one more fine lattice in the future. One can see that the uncertainty of the cutoff effect is dominant in the charm quark sector. For the total contribution of $[a_\mu^{\rm hvp}]^l_{\rm lat}+[a_\mu^{\rm hvp}]^s_{\rm lat}+[a_\mu^{\rm hvp}]^c_{\rm lat}$, the uncertainty in the light quark sector is still dominant. 

\subsection{Analysis of momentum-space integration scheme on 128$^4$ lattice}\label{seq:mom_rep}
Compared to the coordinate-space integration scheme, the momentum-space integration scheme is rather straightforward for performing the integral in Eq.~(\ref{eq:a_mu_infinite}) once $\Pi(0)$ is determined by the zero-momentum extrapolation of VPF (see Eq.~(\ref{eq:a_mu_mom})). The advantage in this work over prior ones \cite{DellaMorte:2011aa,Aubin:2012me,Aubin:2013daa,Burger:2013jya,DellaMorte:2017dyu} is that we can not only access VPF in the low-momentum region but also have high resolution in terms of $Q^2$ without resort to the twisted boundary condition for the valence quark by using a large lattice size larger than (10 fm$)^4$. Our large lattice is also useful for reducing the uncertainty due to zero-momentum extrapolation and does not introduce the partially quenching effect for the different boundary conditions between the sea and valence quarks.

\begin{figure}
  \begin{center}
    \includegraphics[width=100mm]{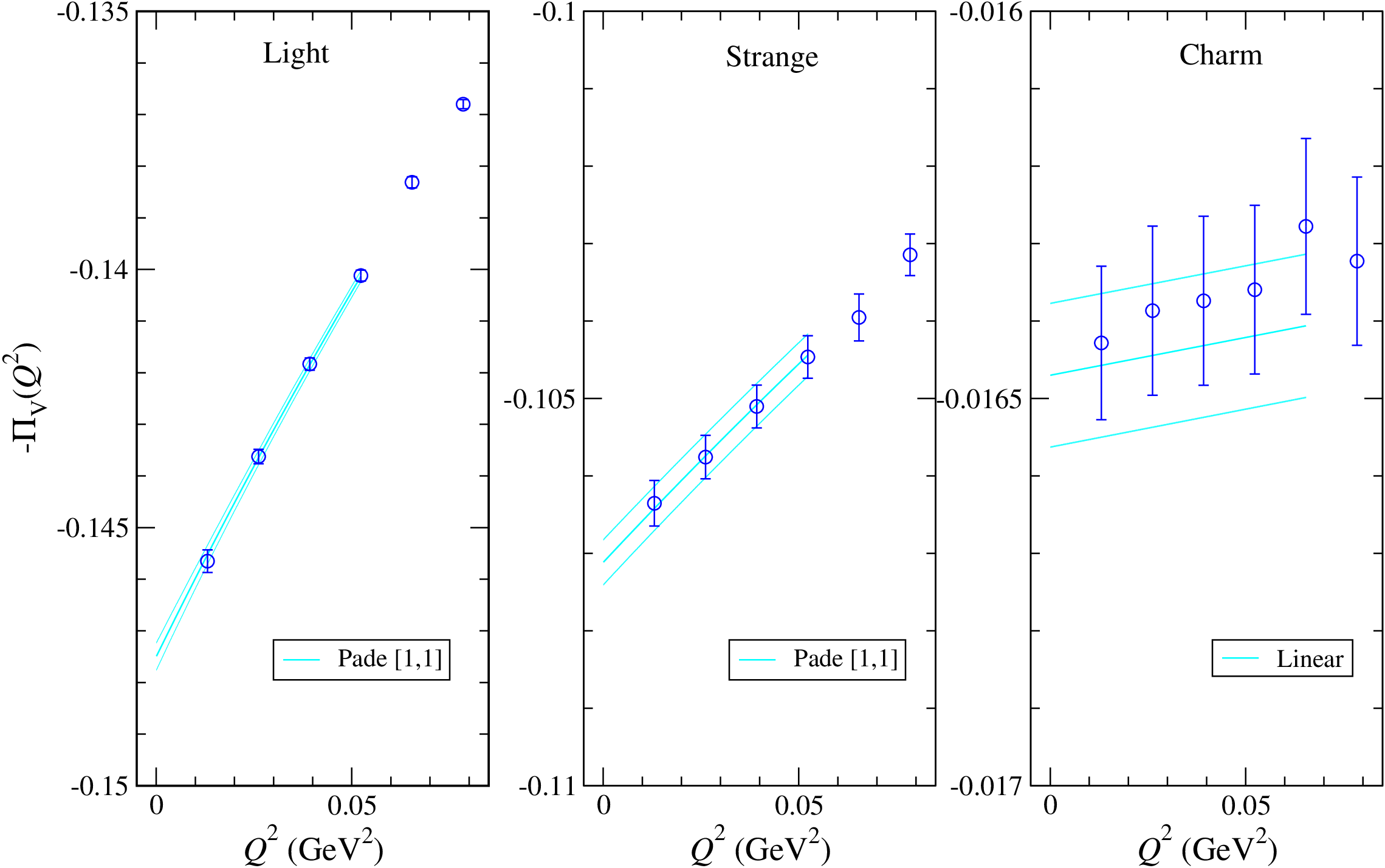}
    \caption{$Q^2$ dependence of VPF in light (left), strange (middle) and charm (right) quark sectors. Solid lines denote the fit results including the statistical error with Pad\'e [1,1] approximation (left and middle) and linear function (right).}\label{fig:vp_q2dep}
  \end{center}
\end{figure}

\begin{figure}
  \begin{center}
    \includegraphics[width=75mm]{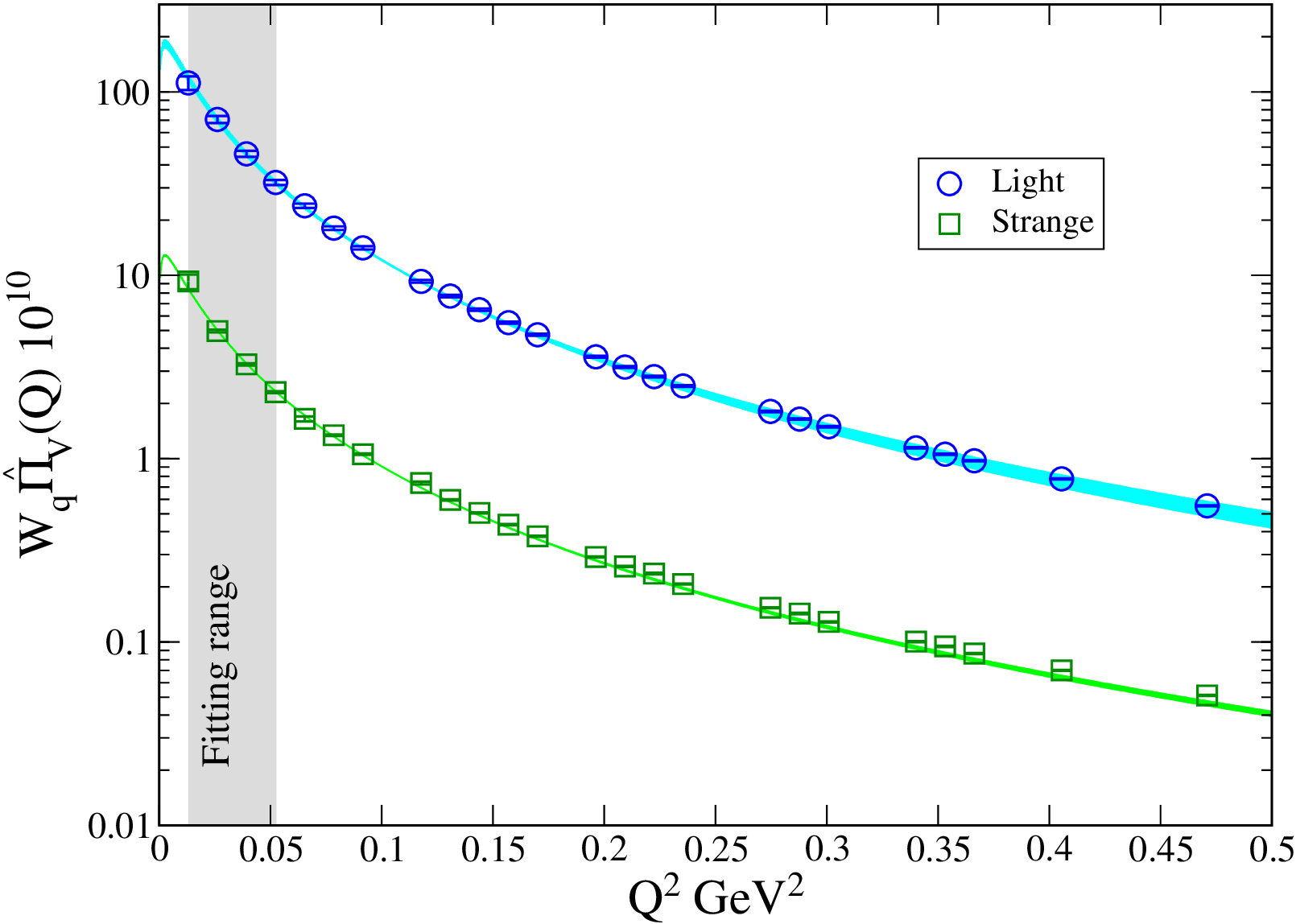}
    \hspace{3mm}
    \includegraphics[width=75mm]{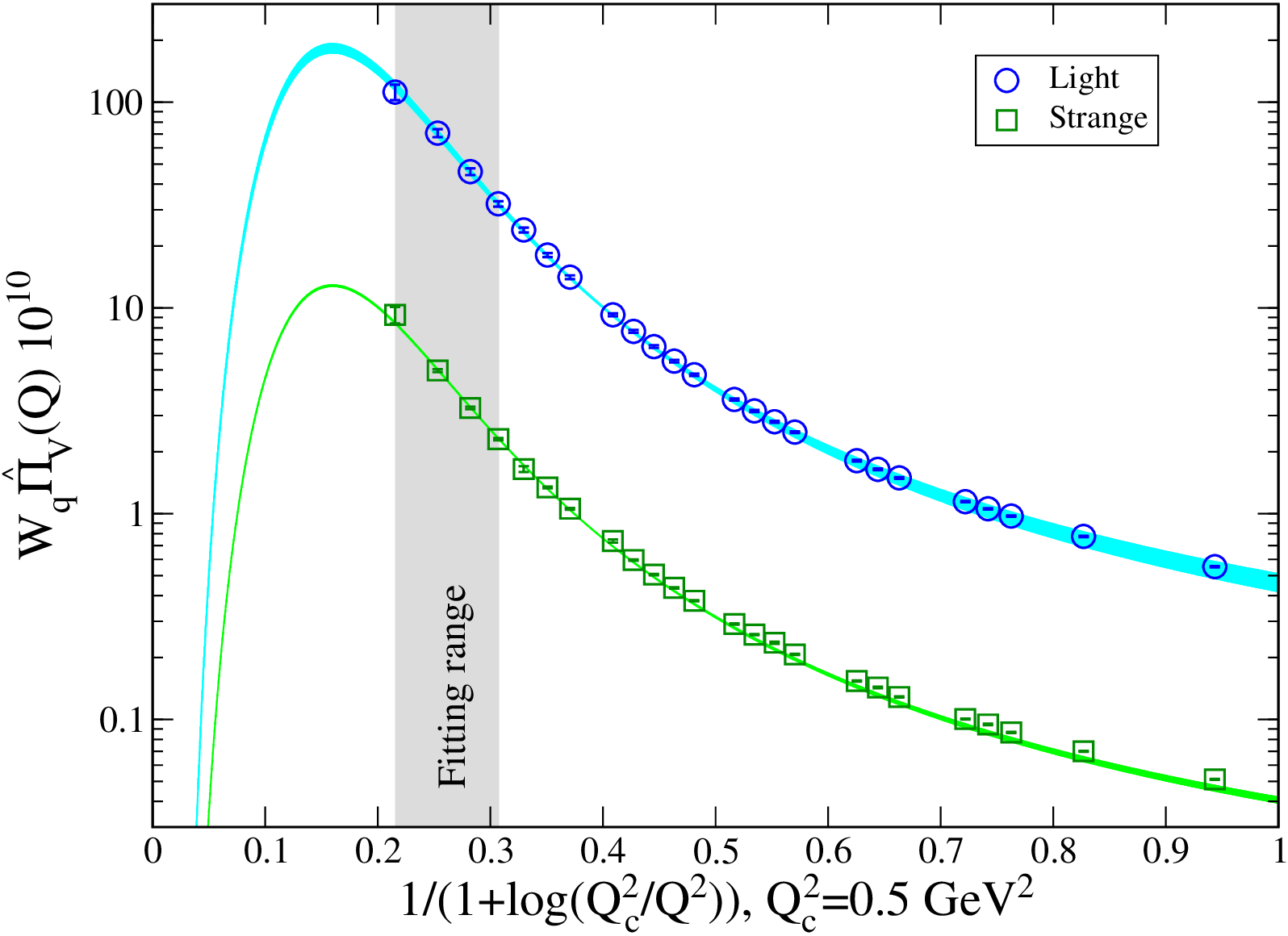}
    \caption{$Q^2$ dependence of integrand in Eq.~(\ref{eq:a_mu_infinite}) up to $Q^2=0.5$ GeV$^2$ at light and strange quark sectors. The horizontal axis is rescaled to a dimensionless quantity $1/(1+\ln(Q_c^2/Q^2))$~\cite{Boyle:2011hu} with $Q_c^2=0.5$ GeV$^2$ in the right panel. Curved bands show the fit results including the statistical error. The shaded vertical band denotes the fitting range for $\Pi(Q)$.}\label{fig:amu_q2dep}
  \end{center}
\end{figure}

We plot the LQCD data of VPF in each quark sector in Fig.~\ref{fig:vp_q2dep}, where the fit results of Pad\'e approximation of order [1,1] for the light and strange quark sectors using  $Q_{\rm fit}^2\le 0.05$ GeV$^2$ and linear function for charm quark using $Q^2_{\rm fit}\le 0.065$ GeV$^2$ are also shown. We observe that VPF in the light and strange quark sectors have stronger slope around the zero momentum region than that in the charm quark sector. As shown in Fig.~\ref{fig:vp_q2dep}, the Pad\'e approximation of lowest order of $[1,1]$ well describes such a lattice data with reasonable $\chi^2$/dof$<1$ in the correlated fit. Although we employ a narrow fit range close to zero-momentum, Fig.~\ref{fig:amu_q2dep} shows that the fit function agrees with the $Q^2$ dependence of LQCD data up to $Q^2=$0.4 GeV$^2$, which is far beyond the fitting range. This behavior indicates the lowest Pad\'e approximation, which consists of single pole dominance, is a reasonable approximation in the IR regime. Since VPF multiplied by the weight function $W_q$ in $Q^2\ge 0.5$ GeV$^2$ gives tiny contribution to the total $a_\mu^{\rm hvp}$ as mentioned in Sec.~\ref{subsec:momentum-integration}, we evaluate the integral without pQCD part (the third term of Eq.~(\ref{eq:a_mu_mom})) in our analysis. In fact, the LQCD data of integral larger than $Q^2\approx0.5$ GeV$^2$ are below 0.5$\times 10^{-10}$ corresponding to less than 0.1\% for $a_\mu^{\rm hvp}$ (see the right panel of Fig.~\ref{fig:amu_q2dep}), and it is then negligible. Therefore we hereafter estimate $a_\mu^{\rm hvp}$ using integrals up to $Q^2=0.5$ GeV$^2$. 

\begin{figure}
  \begin{center}
    \includegraphics[width=120mm]{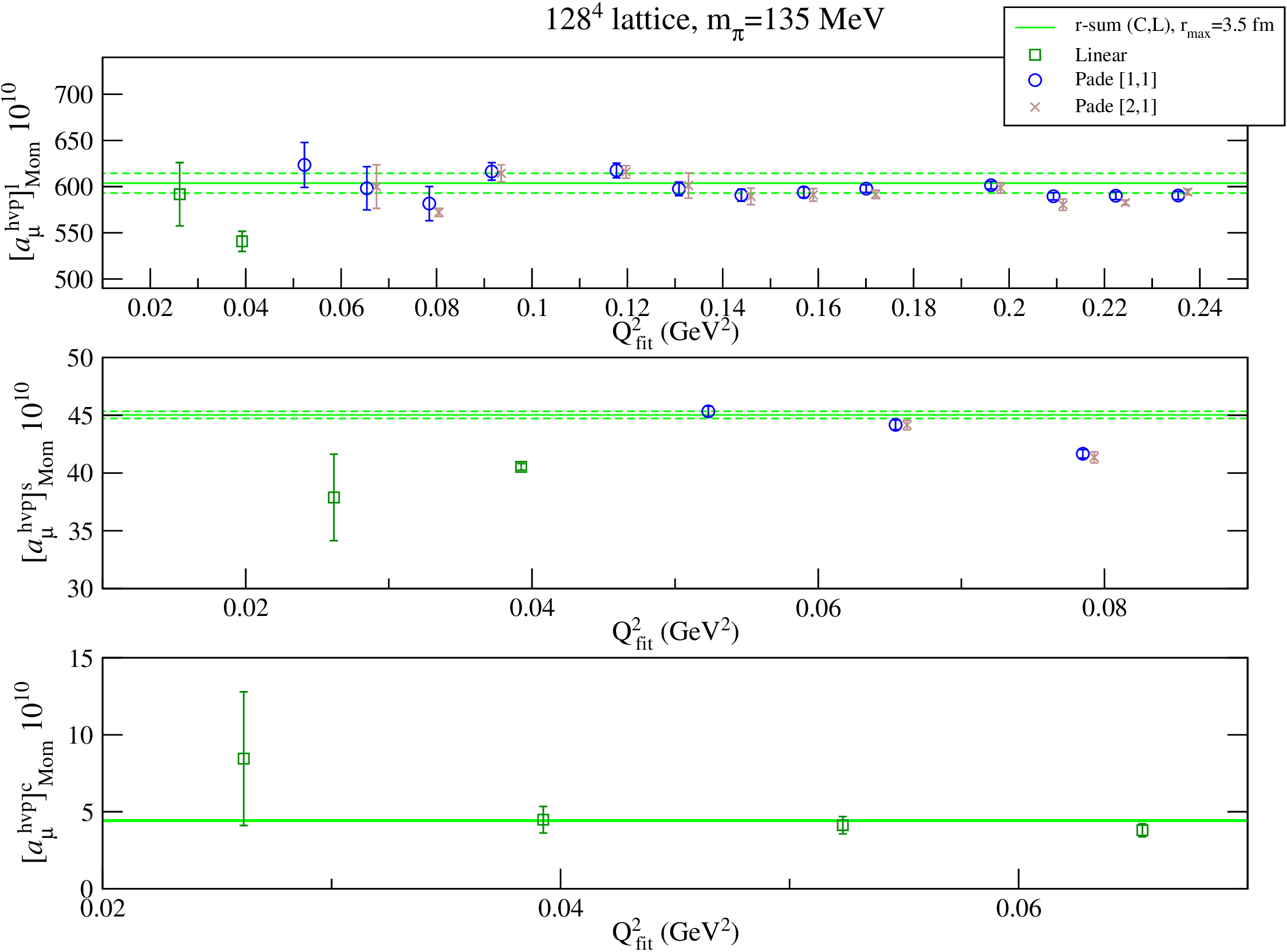}
    \caption{$Q_{\rm fit}^2$ dependence in  zero momentum extrapolation with linear function and Pad\'e approximation for VPF in light (top), strange (middle) and charm (bottom) quark sectors.   Horizontal lines are the central value (solid) and statistical error (dashed) of the result in coordinate-space integration scheme with $r_{\rm cut}\approx 3.5$ fm.}\label{fig:amu_fitdep}
  \end{center}
\end{figure}

Since the integrand has the sharp peak structure much below the minimum momentum squared $Q_{\rm min}^2\approx 0.013$ GeV$^2$ allowed on our ensemble (see Fig.~\ref{fig:amu_q2dep}), the integral in Eq.~(\ref{eq:a_mu_mom}) is sensitive to the extrapolation procedure from $Q^2_{\rm min}$ to zero. We employ the linear extrapolation and the Pad\'e approximation of order [1,1] and [2,1]. Figure~\ref{fig:amu_fitdep} compares the results of $[a_\mu^{\rm hvp}]_{\rm Mom}$ obtained by both extrapolation methods with varying the fitting ranges from $Q^2_{\rm min}$ to $Q_{\rm fit}^2$. In case of the linear extrapolation the results of $[a_\mu^{\rm hvp}]_{\rm Mom}$ show significant $Q_{\rm fit}^2$ dependence, due to the higher order term than $\mathcal O(Q^2)$ even in $Q^2\approx 0.013$ GeV$^2$, except for the charm quark sector. On the other hand, we observe little $Q_{\rm fit}^2$ dependence for $[a_\mu^{\rm hvp}]_{\rm Mom}$ with the Pad\'e approximation of order [1,1] and [2,1] up to $Q_{\rm fit}^2=0.235$ GeV$^2$ in the light quark sector. We find that, even with different $Q_{\rm fit}^2$ and different orders of Pad\'e approximation,  the result in momentum-space integration scheme is in good agreement with $[a_\mu^{\rm hvp}]^{\rm CL}_{\rm lat}(r_{\rm cut})$ at $r_{\rm cut}=3.5$ fm within 1.5 $\sigma$ error, and this is thus consistency test of the scheme independence. We also find that the systematic uncertainty due to fitting with Pad\'e approximation is negligible in our study on $L=10.8$ fm lattice.

Here we notice the strong $Q_{\rm fit}^2$ dependence for the results in the strange quark sector appears in Fig.~\ref{fig:amu_fitdep}. In this case, an extra lattice cutoff effect of $\mathcal O(am_{\rm s}(aQ)^2)$, which is not described by the naive Pad\'e approximation, may arise in strange quark sector. More detailed study will be needed in the future. 

\begin{figure}
  \begin{center}
    \includegraphics[width=120mm]{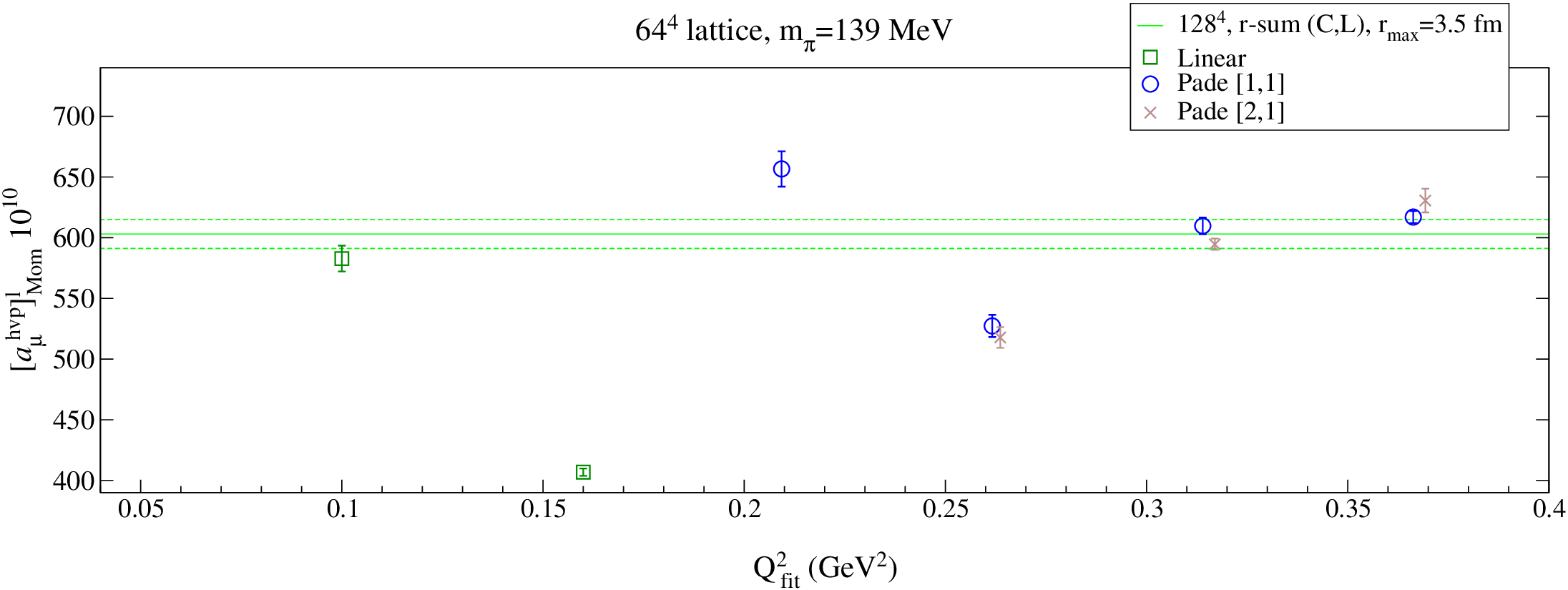}
    \caption{Same as Fig.~\ref{fig:amu_fitdep} on 64$^4$ lattice.}\label{fig:amu_fitdep_64}
  \end{center}
\end{figure}

In contrast to the $128^4$ lattice, Fig.~\ref{fig:amu_fitdep_64} shows the significant $Q_{\rm fit}^2$ dependence for the results with both extrapolation methods on the $64^4$ lattice since the low $Q^2$ data has coarse resolution on this lattice. Our LQCD study suggests that the lattice size with $L=5.4$ fm at the physical pion mass, corresponding to $m_\pi L=3.8$, is not large enough for the momentum-space integration scheme to obtain a reliable result of $a_\mu^{\rm hvp}$ because of large FV correction. 

We remark that the statistical precision of the result for $[a_\mu^{\rm hvp}]_{\rm Mom}$ is more easily obtained than that for $[a_\mu^{\rm hvp}]_{\rm lat}$. This is because of a noise cancellation in $\hat\Pi(Q)$ between the extrapolated $\Pi(0)$ and $\Pi(Q)$, which are highly correlated with each other. In addition, in the momentum-space integration scheme, we do not need to introduce the truncation of the integration range corresponding to the IR truncation $r_{\rm cut}$ in the coordinate-space integration scheme. This indicates the possibility that once we have low $Q^2$ data covering a peak position of $W_q$ ($Q^2\sim 0.003$ GeV$^2$), for which we need to prepare a box size 2 times larger than in this study, we can obtain a high precision result with smaller statistical and systematic errors than the coordinate-space integration scheme.

\subsection{Discussion}\label{sec:discussion}
We obtain the connected $a_\mu^{\rm hvp}$ in the light, strange and charm quark sectors at the physical point: 
\begin{eqnarray}
  a_\mu^{\rm hvp} = \left\{ \begin{array}{cc}
    673(9)(11)\times 10^{-10} & \textrm{[light]}\\
    52.1(2)(5)\times 10^{-10} & \textrm{[strange]}\\
    11.7(2)(1.6)\times 10^{-10} & \textrm{[charm]}\\
  \end{array}
  \right.,\label{eq:amu_uds}
\end{eqnarray}
where the first error is statistical for $(\Gamma,\Gamma')=$(L,L) with the constant fit and the second one is systematic for the uncertainty in the continuum extrapolation explained in Sec.~\ref{sec:latart}. We find that the statistical and systematic errors for the light quark sector gives the leading contribution to the total error. The contributions from the strange and charm quark sectors are minor effects. 

Here we make two remarks:
\begin{enumerate}
\item Our choice of $r_{\rm cut}\approx 3.5$ fm in the coordinate-space integration scheme, which is larger than 3 fm value employed in Refs.~\cite{Borsanyi:2017zdw,Blum:2018mom}, is large enough to control the IR truncation.  In Figs.~\ref{fig:amu_tdep_local} and \ref{fig:amu_local} we observe that the integrand has a nonzero value of $23(10)\times10^{-10}$ at $r_{\rm cut}\approx 3$ fm in the $(\Gamma,\Gamma')=$(L,L) channel on the 128$^4$ lattice and the integral is still increasing, while the integrand is consistent with zero at $r_{\rm cut}\approx 3.5$ fm and the integral does not depend on $r_{\rm cut}$ even if we use larger $r_{\rm cut}$. High precision data on a lattice larger than (10 fm$)^4$ at the physical point allow us to evaluate the integral with the IR truncation effect under control.
\item The scaling properties presented in Sec.~\ref{sec:latart} are similar to the domain-wall fermion case~\cite{Blum:2018mom}, though the computational cost is much lower for Wilson-type quark action. The continuum extrapolation is straightforward and theoretically robust for Wilson-type quark action compared to the staggered fermion case~\cite{Borsanyi:2016lpl,Borsanyi:2017zdw}. 
\end{enumerate}
 
\begin{figure}
  \begin{center}
    \includegraphics[width=73mm]{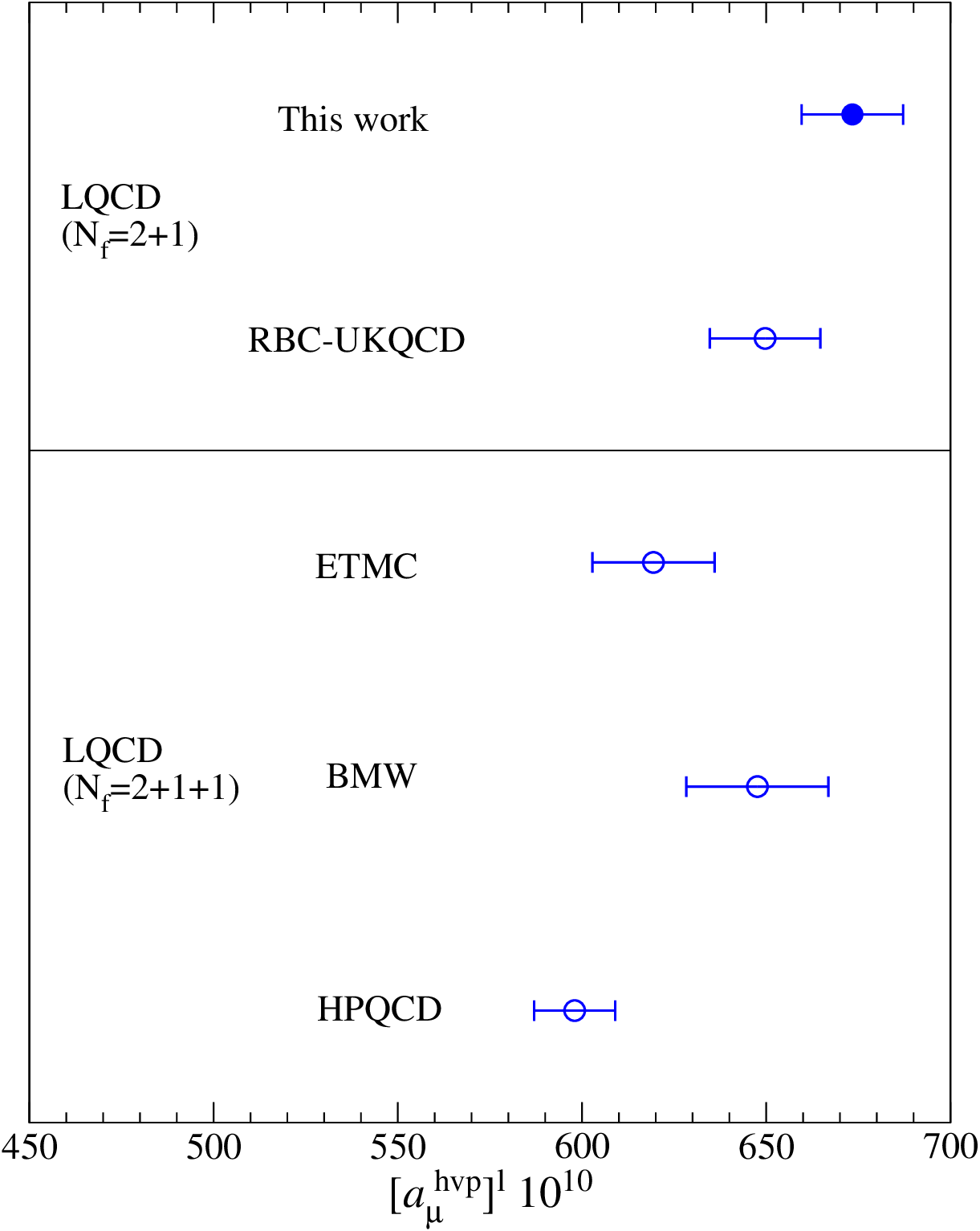}
    \hspace{3mm}
    \includegraphics[width=71mm]{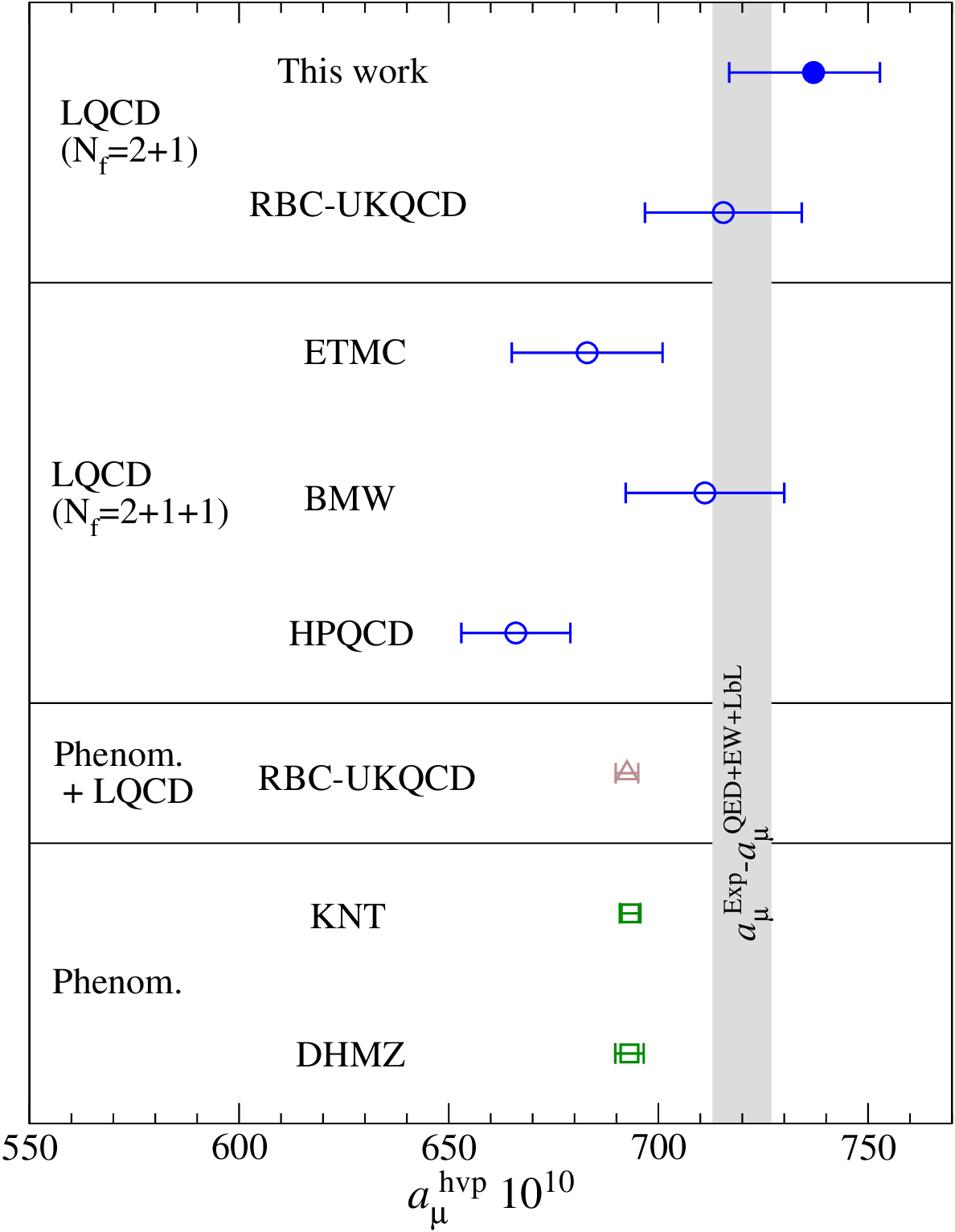}
    \caption{Summary plot of the connected $a_\mu^{\rm hvp}$ in the light quark sector $[a_\mu^{\rm hvp}]^l$ (left) and the full result of $a_{\mu}^{\rm hvp}$ (right) in comparison with recent LQCD results ($N_f\ge 3$) by BMW~\cite{Borsanyi:2017zdw}, ETMC~\cite{Giusti:2018mdh}, HPQCD~\cite{Chakraborty:2016mwy}, RBC-UKQCD~\cite{Blum:2018mom} collaborations and phenomenological estimate obtained with the experimental R-ratio by DHMZ~\cite{Davier:2017zfy} and KNT~\cite{Keshavarzi:2018mgv}. Shaded vertical band shows the ``experimental'' $a_\mu^{\rm hvp}$ estimated by the difference between the BNL experimental value of $a_\mu$ and the theoretical value with QED and EW including the light-by-light scattering contribution. The error bar for $[a_\mu^{\rm hvp}]^l$ in this work represents the combined error with the statistical one and the systematic one due to cutoff effect. Additional uncertainties of missing disconnected diagram and IB effect are included in the error bar of $a_{\mu}^{\rm hvp}$ in this work.}\label{fig:summary}
  \end{center}
\end{figure}

In this paper, we concentrate on the connected HVP diagram, while there are some missing diagrams of the isoscalar contribution with the disconnected diagram and the isospin breaking (IB) term due to the QED correction. Referring to the recent work in Refs.~\cite{Borsanyi:2017zdw,Blum:2018mom}, we conservatively add the systematic error of the quark disconnected diagram contribution as $-$2\% effect and the IB effect as $+$1\% error to the total contribution. We then find that
\begin{equation}
  a_\mu^{\rm hvp} = 737(9)(^{+13}_{-18})\times 10^{-10}, 
\end{equation}
where the first error is statistical and the second one represents the total systematic error obtained in the quadrature. The magnitude of the error is still 2.7\%, in which the systematic error, mainly due to the uncertainty of disconnected diagram, is more than 2 times larger than the statistical one. Compared to other lattice results ($N_f\ge3$) (see Fig.~\ref{fig:summary}), our value is consistent with the results by RBC-UKQCD~\cite{Blum:2018mom} and BMW~\cite{Borsanyi:2017zdw} collaborations, while we find a slight tension with recent results of the ETMC~\cite{Giusti:2018mdh}, HPQCD~\cite{Chakraborty:2016mwy} collaborations, and two-$\sigma$ deviation from the phenomenological estimates~\cite{Davier:2017zfy,Keshavarzi:2018mgv}. Our result seems to favor the ``experimental'' $a_\mu^{\rm hvp}$, which is defined as the difference between the BNL experimental value of $a_\mu$ and the theoretical calculation with QED and EW including the light-by-light scattering contribution in Ref.~\cite{Olive:2016xmw}.

\section{Summary}\label{sec:summary}
We have studied the systematic uncertainties in the LQCD calculation of $a_\mu^{\rm hvp}$ on the PACS10 gauge configurations which have a greater than (10 fm$)^4$ box size at the physical point with two different lattice cutoffs. This study and previous work~\cite{Izubuchi:2018tdd} are the direct LQCD calculations without use of any ansatz or reliance on any effective models. The optimized LQCD calculation of HVP on sufficiently large lattice size at the physical point allows us to access the deep IR regime where the contributions of multihadron states become manifest. Our study points out that such contributions may be larger than the estimate in the leading order of ChPT. In Fig.~\ref{fig:summary} we observe that our result of $a_\mu^{\rm hvp}$ is relatively larger than that of other LQCD studies. The reason for such a tendency may be due to the discrepancy between LQCD and ChPT (or related phenomenological models) including only a two-pion state contribution, which was applied to evaluate the FV correction in other LQCD studies, as discussed in Sec.~\ref{sec:FV}. We have also investigated the lattice cutoff effect in the coordinate-space integration scheme using data at two different cutoffs. We find that the cutoff effect is tamed for the local vector current on our gauge configurations. Furthermore, the momentum-space integration scheme on a $L> 10$ fm lattice yields high quality data for a VPF close to $Q^2=0$, which substantially reduces uncertainty in the zero-momentum extrapolation. With a careful study of the extrapolation procedure dependence we have confirmed the consistency between the results in the momentum- and coordinate-space integration schemes. 

The total error for the result of $a_\mu^{\rm hvp}$ is 2.7\%, in which the statistical error is 1.2\% and the remaining is the systematic uncertainty. We plan to reduce both the statistical and systematic errors with additional calculations, including one finer lattice, disconnected diagram, and QED effect in future. Here we will point out a possibility that the momentum-space integration scheme with $L> 20$ fm covers the peak position of kernel function in low $Q^2$ regime so that it could be a rigorous test for the LQCD scheme. We leave it to future work.

\section*{Acknowledgments}
We would like to thank Taku Izubuchi and Christoph Lehner, and also PACS collaboration members for helpful discussion and support. The computation code was developed based on Columbia Physics System(CPS) incorporating the optimized OpenQCD system\footnote{See http://luscher.web.cern.ch/luscher/openQCD/}. This work is supported in part by the U.S.-Japan Science and Technology Cooperation Program in High Energy Physics for FY2018,  Interdisciplinary Computational Science Program of Center for Computational Sciences (CCS) at the University of Tsukuba, general use No.~G18001 at ACCC, and the HPCI System Research project (Project ID:hp180126, hp190068). Numerical calculations were performed on the K computer in RIKEN Center for Computational Science (R-CCS), Hokusai at Advanced Center for Computing and Communication (ACCC) in RIKEN, XC40 at YITP in Kyoto University, the Fujitsu PRIMERGY CX600M1/CX1640M1 (Oakforest-PACS) in the Information Technology Center, The University of Tokyo, and the computer facilities at Research Institute for Information Technology, Kyushu University. 

\appendix
\section{The derivation of coordinate representation}\label{appendix:dev}
Fourier transformation of Eq.~(\ref{eq:vvx}) is defined as
\begin{eqnarray}
  G(Q) = \int d^4xe^{iQx} C(x) = 3Q^2\Pi(Q),\label{eq:vvq}.
\end{eqnarray}
$\hat\Pi$ in Eq.~(\ref{eq:Pihat}) can be represented as
\begin{eqnarray}
  \hat\Pi(\omega) = \frac{1}{3\omega^2}G(\omega) - \frac{1}{3\omega^2}G(\omega)\Big|_{\omega=0},
\end{eqnarray}
where the second term is expanded as
\begin{equation}
  \frac{1}{3\omega^2}G(\omega)\Big|_{\omega=0} = \frac{1}{3\omega^2}\Big[ G(0) + \frac{1}{2}\omega^2G^{\prime\prime}(0)+\mathcal O(\omega^4)\Big]
\end{equation}
and we obtain
\begin{eqnarray}
  \hat\Pi(\omega) =  \frac{1}{3\omega^2}G(\omega) - \frac{1}{3\omega^2}G(0) - \frac 1 6 G^{\prime\prime}(0).
\end{eqnarray}
In general, the second derivative of $G$ with respect to $\omega=|Q|$ can be expressed as
\begin{eqnarray}
  G^{\prime\prime}(\omega) &=& \int d^4xe^{iQx}\Big[
    \sum_{\mu\nu}(-x_\mu x_\nu)\frac{\omega^2}{Q_\mu Q_\nu}e^{iQx}C(x)\Big|_{Q_\mu\ne0,Q_\nu\ne0}\Big]_{\omega=|Q|},
\end{eqnarray}
where the terms with odd power of $x_\mu$ vanish in the coordinate integral. By substituting the above equation into Eq.~(\ref{eq:a_mu_infinite}), we can obtain Eq.~(\ref{eq:a_mu_coor_gen}).

\bibliographystyle{apsrev4-1}
\bibliography{ref}
\end{document}